\documentclass[nofootinbib,aps,11pt]{revtex4-1}
\usepackage{amsmath, fullpage}
\usepackage{booktabs} 
\usepackage{graphicx}
\graphicspath{{./}{../}}
\usepackage{array}
\usepackage{hyperref}
\usepackage{cancel}
\usepackage{amssymb}
\usepackage{textcomp}
\usepackage{amsmath}
\usepackage{bm}
\usepackage{times}
\usepackage{epsfig}
\usepackage{color}
\usepackage{float}
\usepackage{comment}
\usepackage{tikz}
\usepackage{booktabs}
\usepackage{graphicx,color,dcolumn,booktabs,bm}
\usepackage{adjustbox}
\usepackage{amssymb,graphicx}
\usepackage{pifont}

\begin{document}

\title{Inelastic Kinetic Decoupling in WIMP Leptogenesis: A Minimal $Z_4$ Model}
\author{An Liu}
\email{liuan7@mail2.sysu.edu.cn}
\author{Hong-Hao Zhang}
\email[Corresponding author. ]{zhh98@mail.sysu.edu.cn}
\affiliation{School of Physics, Sun Yat-Sen University, Guangzhou 510275, China}
\begin{abstract}
We study a minimal $Z_4$-symmetric extension of the Standard Model in which a singlet Dirac fermion $\chi$ constitutes dark matter, whose annihilations generate a lepton asymmetry through WIMP leptogenesis. Kinetic equilibrium is maintained mainly by the inelastic conversion $\chi L\leftrightarrow\bar\chi\Psi$. Compared with the near elastic regime, a sizable mass gap suppresses the conversion rate more rapidly as the Universe cools. This naturally drives kinetic decoupling to earlier times, even away from resonance. Starting from the full inelastic collision operator, we derive an effective momentum transfer rate for kinetic decoupling. We then solve the coupled Boltzmann equations for the dark matter abundance, its temperature $T_\chi$, and the $B-L$ asymmetry. Early kinetic decoupling can significantly modify the relic abundance, although near resonance this modification can be reduced by annihilation-induced heating. The $B-L$ asymmetry can be affected when asymmetry production overlaps with a substantial departure from kinetic equilibrium. In our scan, the largest deviations occur mainly for $x_{\rm lep}\gtrsim x_{\rm cd}$. After imposing our constraints, asymmetry production typically ends before chemical decoupling, leaving the final asymmetry essentially unchanged. For indirect detection, we generate the photon spectrum with Pythia and evaluate the Fermi-LAT dwarf spheroidal likelihood.

\end{abstract}

\maketitle
\thispagestyle{empty}

\section{Introduction}

The Standard Model (SM) of particle physics has achieved remarkable success in
describing a wide range of phenomena probed by collider experiments and
precision tests. Nevertheless, it leaves several fundamental questions
unanswered, including the particle nature of dark matter (DM) and
the origin of the cosmic baryon asymmetry
	\cite{Bertone:2004pz,Planck:2018vyg}. 

Astrophysical and cosmological observations, including galaxy rotation curves,
gravitational lensing, cosmic microwave background (CMB) anisotropies, and
the observed large-scale distribution of matter, provide compelling evidence for the existence
of dark matter. The measured relic abundance is
\begin{equation}
\Omega_{\rm DM}h^2\simeq 0.12 ,
\end{equation}
while the microscopic identity of dark matter remains unknown. Meanwhile, the
observed baryon asymmetry is~\cite{Planck:2018vyg,Fields:2019pfx}
\begin{equation}
\eta_B=\frac{n_B-n_{\bar B}}{n_\gamma}\simeq 6.1\times10^{-10}.
\end{equation}
Within the Standard Model, electroweak baryogenesis cannot account for the observed baryon asymmetry because the available CP violation is insufficient and the electroweak transition is a smooth crossover rather than a strongly first-order phase transition~\cite{DOnofrio:2014rug}.
This motivates alternative baryogenesis mechanisms, including leptogenesis, in which a lepton asymmetry is partially converted into a baryon asymmetry by electroweak sphaleron processes~\cite{Fukugita:1986hr,Buchmuller:2005eh,Manton:1983nd,Klinkhamer:1984di,Khlebnikov:1988sr}.
These two cosmological puzzles
therefore provide strong motivations for physics beyond the Standard Model
(BSM) \cite{Elor:2018twp,Shuve:2020evk,Hall:2019rld,Carr:2019hud,Ma:2006fn,Falkowski:2017uya,Liu:2020mxj}.

Interestingly, dark matter and baryogenesis may share a common origin
\cite{Petraki:2013wwa,Zurek:2013wia,Kaplan:2009ag}. A
variety of BSM frameworks have explored scenarios in which the dark sector not
only accounts for the observed relic abundance but also contributes to the
generation of the cosmic baryon asymmetry
\cite{Cui:2011ab,Bernal:2013bga}.
Although conventional thermal leptogenesis with hierarchical heavy Majorana neutrinos is
theoretically appealing, the high mass scale typically required makes direct experimental
tests difficult
\cite{Buchmuller:2005eh,Davidson:2002qv}.

This motivates realizations of leptogenesis at lower scales
\cite{Calabrese:2023bxz}. Conversion-driven leptogenesis offers another connection between dark matter freeze-out and baryogenesis, with CP-violating decays and inverse decays linking dark matter to a charged mediator~\cite{Heisig:2024conversion}. In particular, WIMP
leptogenesis establishes a direct connection between dark matter annihilation
and baryogenesis, in which dark matter annihilations can simultaneously determine
the relic abundance and generate a lepton asymmetry through CP-violating processes
\cite{Cui:2011ab,Bernal:2012gv,Mahanta:2022WIMPy}. However, most studies assume that dark
matter remains in kinetic equilibrium with the thermal bath during chemical
decoupling,
\begin{equation}
T_\chi=T ,
\end{equation}
so that the conventional Boltzmann equation for the number density is sufficient.
This assumption, however, need not hold. Early kinetic decoupling (eKD) can
occur when momentum exchange becomes inefficient before chemical freeze-out.
Its impact on the relic abundance can be particularly pronounced in scenarios
involving resonant annihilation, Sommerfeld enhancement, or forbidden annihilation channels
\cite{Binder:2017rgn,Duch:2017nbe,Abe:2020obo,Bi:2011qm,Binder:2021bmg,Liu:2023kat,Aboubrahim:2023yag}. In such cases,
\begin{equation}
T_\chi\neq T .
\end{equation}

Although early kinetic decoupling has been extensively studied in calculations of the dark matter
relic abundance~\cite{Bringmann:2006mu,Binder:2017rgn}, its impact on lepton asymmetry generation
in WIMPy leptogenesis has received relatively little attention. In this work, we
investigate this issue within a minimal $Z_4$-symmetric extension of the SM.
The model contains a singlet Dirac fermion $\chi$ as the dark matter candidate,
a vectorlike fermion doublet $\Psi$, and pseudoscalar mediators. The key
feature is that kinetic equilibrium is maintained mainly through the inelastic
conversion process
\begin{equation}
\chi L\leftrightarrow \bar\chi\Psi .
\end{equation}
As the production of the heavier state $\Psi$ becomes increasingly Boltzmann suppressed,
this conversion process naturally becomes inefficient and induces early kinetic
decoupling.

In this work, we formulate the full inelastic collision operator and
define an effective momentum-transfer rate to
determine the kinetic decoupling temperature. We solve coupled Boltzmann
equations for the dark matter abundance, its kinetic temperature, and the
lepton asymmetry. The visible sector $B-L$ asymmetry is then obtained using
spectator relations. We find that eKD can substantially modify the relic abundance
when the annihilation cross section has strong velocity dependence, while the
final asymmetry is generally less sensitive. In viable parameter regions, the
ratio between the relic abundance obtained with the coupled treatment and the
standard treatment can deviate significantly from unity, whereas the generated
$B-L$ asymmetry remains comparatively stable. We further investigate the corresponding indirect detection signals.

This paper is organized as follows. In Sec.~\ref{sec:model}, we introduce
the particle content and interactions of the $Z_4$ model. In Sec.~\ref{sec:annihilations}, we study
dark matter annihilation, kinetic decoupling, and lepton asymmetry generation.
The numerical analysis and phenomenological implications are
presented in Sec.~\ref{sec:pheno}. Indirect detection is discussed in Sec.~\ref{sec:indirect}. Finally, Sec.~\ref{sec:conclusion} summarizes our conclusions.

\section{Model}
\label{sec:model}

We consider a $Z_4$ model based on the WIMPy leptogenesis framework of Ref.~\cite{Cui:2011ab}. The model extends the Standard Model by introducing two real gauge-singlet pseudoscalar mediators $\phi_1$ and $\phi_2$, a light right-handed Weyl fermion $n_R$, a vectorlike fermion doublet $\Psi_{L,R}=(\psi^0,\psi^-)^T_{L,R}$, and a gauge-singlet Dirac fermion $\chi$ as the dark matter candidate. The particle content and corresponding symmetry assignments are summarized in Table~\ref{tab:Z4}. The unbroken $\mathbb{Z}_4$ symmetry contains a $\mathbb{Z}_2$ subgroup under which $\chi$ is odd and all other fields are even, ensuring dark matter stability.
We work in the scalar mass eigenstate basis, with the two mediator masses denoted by $m_{\phi_1}$ and $m_{\phi_2}$.
We focus on the WIMP regime, where the dark matter relic abundance is determined by thermal freeze-out.

\begin{table}[h]
\centering
\renewcommand{\arraystretch}{1.1}
\setlength{\tabcolsep}{6pt} 
\begin{tabular}{@{}lcccc@{}}
\toprule
\hline\hline
& $\rm SU(2)_L$ & $\rm U(1)_Y$ & $\mathbb{Z}_4$ & $U(1)_F$ \\
\hline
$\chi$ & 1 & 0 & $i$ & 0 \\
$\Psi_{L,R}$ & 2 & $-\frac{1}{2}$ & $-1$ & $+1$ \\
$n_R$ & 1 & 0 & $-1$ & $+1$ \\
$\phi_{1,2}$ & 1 & 0 & $-1$ & 0 \\
$L$ & 2 & $-\frac{1}{2}$ & 1 & $+1$ \\
$H$ & 2 & $+\frac{1}{2}$ & 1 & 0 \\
\hline\hline
\bottomrule
\end{tabular}
\caption{Particle content and charge assignments under $SU(2)_L \times U(1)_Y \times \mathbb{Z}_4$, together with the global $U(1)_F$ charges.}
\label{tab:Z4}
\end{table}

The relevant fermion mass terms are
\begin{align}
-\mathcal{L}_{\rm mass}\supset m_\chi\bar\chi\chi
+\left(m_\Psi\bar\Psi_L\Psi_R+\text{h.c.}\right),
\end{align}
The relevant Yukawa interactions are
\begin{align}
-\mathcal{L}_{\rm Y}  \supset  & \nonumber
    \frac{i}{2} y_{\chi\alpha} \phi_\alpha \bar{\chi^c} \gamma_5 \chi+  i \, \lambda_{L\alpha } \phi_\alpha  \bar{L}
 {\Psi_R} + \lambda_{n} \bar{\Psi}_L \tilde{H} n_R +\text{h.c.},
\end{align}
The dark matter Yukawa interaction corresponds to a particular pseudoscalar coupling structure within the more general framework of Ref.~\cite{Cui:2011ab}.
Here, $L$ denotes the Standard Model lepton doublet, and $\widetilde H=i\sigma_2H^*$ is the conjugate Higgs doublet. The index \( \alpha = 1,2 \) labels the two pseudoscalar fields. Pseudoscalar mediators are preferred over scalars, as they avoid the velocity suppression of the dark matter annihilation cross section \( \chi\chi\to L\bar\Psi \), thereby facilitating efficient leptogenesis~\cite{Cui:2011ab}. 
In order to satisfy the Sakharov conditions~\cite{Sakharov:1967dj}, the dark matter annihilation process must violate CP symmetry. CP violation requires physical relative phases among the Yukawa couplings that cannot be removed by field redefinitions.
To ensure gauge anomaly cancellation, the fermion doublet $\Psi$ is vectorlike under $SU(2)_L$~\cite{Witten:1982fp}. The singlet $n_R$ is a massless hidden Weyl fermion and does not constitute a dark matter component. At the level of the perturbative Lagrangian, we impose a global fermion number with charges $Q(L)=Q(e_R)=Q(\Psi)=Q(n_R)=1$ and $Q(H)=Q(\phi_\alpha)=Q(\chi)=0$. The SM quarks carry zero charge under this symmetry. The symmetry preserves the Yukawa interactions above and forbids the Majorana mass term for $n_R$. Accordingly, we set $m_{n_R}=0$. Electroweak sphaleron processes violate this fermion number but conserve the total $B-F$ charge. Possible higher dimensional interactions that violate this global fermion number are outside the scope of this work. The coupling $\lambda_n\overline{\Psi_L}\widetilde H n_R$ transfers the asymmetry stored in the vectorlike doublet to the hidden sector through $\Psi\to H n_R$. We work in the regime $\Gamma_{\Psi\to H n_R}\gg H$ during leptogenesis, ensuring efficient asymmetry transfer to the hidden sector. Meanwhile, electroweak gauge interactions are assumed to maintain the vectorlike doublet $\Psi$ in kinetic and chemical equilibrium with the plasma over the temperature range relevant for leptogenesis.

The massless singlet $n_R$ behaves as a relativistic spectator that receives the asymmetry transferred from $\Psi$. While inverse decays $H n_R\to\Psi$ are efficient, they keep $n_R$ in thermal contact with the plasma, with $\Gamma_{H n_R\to\Psi}(T)=\gamma_D(T)/n_{n_R}^{\rm eq}(T)$. Here $\gamma_D$ denotes the equilibrium decay reaction density for $\Psi\to H n_R$, including the relevant charge-conjugate channels. As the $\Psi$ abundance becomes Boltzmann suppressed, this rate rapidly decreases, and we estimate the decoupling temperature from $\Gamma_{H n_R\to\Psi}(T_{n,\rm dec})\simeq H(T_{n,\rm dec})$. After decoupling, its thermal contribution to dark radiation is given by the standard entropy-dilution relation~\cite{Brust:2013xla},
\begin{equation}
\Delta N_{\rm eff}^{\rm th}=\left[\frac{10.75}{g_{*s}(T_{n,\rm dec})}\right]^{4/3}.
\label{eq:delta-neff-thermal}
\end{equation}

The additional relativistic energy density also modifies the expansion rate during radiation domination~\cite{Steigman:2012ve}. We quantify this effect through
\begin{equation}
\frac{\delta H}{H}=\sqrt{1+\frac{\rho_n}{\rho_{\rm rad}}}-1,
\label{eq:delta-h-expansion}
\end{equation}
Here $\rho_{\rm rad}$ excludes the $n_R$ contribution, while $\rho_n$ includes both its thermal component and the energy injected after decoupling. For $\lambda_n=10^{-6}$, the maximum correction over the evolution considered remains at roughly the 1\% level. Below the electroweak crossover, the $\lambda_n$ interaction induces a small mixing between $n_R$ and the neutral component of $\Psi$, opening the light conversion channels $\chi\nu\leftrightarrow\bar\chi n_R$ and $\bar\chi\bar\nu\leftrightarrow\chi\bar n_R$. For $\lambda_n=10^{-6}$, the mixing angle remains below $10^{-6}$, and the corresponding momentum transfer rate satisfies $\gamma_{\rm light}/\widetilde H\ll1$ near kinetic decoupling throughout the parameter region considered. In the selected parameter region, momentum exchange is dominated by inelastic conversion, while both mediator induced elastic scattering and the light conversion channels are subdominant near kinetic decoupling.

To simplify the parameter space considered in this work, we adopt the same assumptions as in Ref.~\cite{Cui:2011ab}. Dark matter annihilation is taken to proceed predominantly through interactions with a single lepton flavor mediated by the lighter pseudoscalar mediator $\phi_1$, while the couplings to other flavors through $\phi_2$ are neglected. The washout processes are likewise assumed to be dominated by $\phi_1$ mediated channels. We denote the lighter mediator $\phi_1$ simply by $\phi$ and neglect the tree-level contribution of $\phi_2$ to annihilation and washout, retaining its contribution to the loop-induced CP asymmetry. 
Although the decay $\phi\to L\bar\Psi$ is kinematically allowed, its decay and inverse decay rates are much faster than the Hubble expansion during leptogenesis, keeping $\phi$ close to chemical equilibrium. We therefore use its equilibrium abundance without evolving a separate Boltzmann equation. Following Ref.~\cite{Cui:2011ab}, we define $\epsilon_\phi$ as the difference between the decay rates of $\phi\to L\bar\Psi$ and $\phi\to\bar L\Psi$, normalized to their sum. Our linear-response estimate with $|\epsilon_\phi|\leq1$ gives a contribution below $1\%$ of the observed asymmetry at sphaleron freeze-out for the selected points, allowing us to neglect this source in Eq.~\eqref{eq:BE_asymmetry}.

\section{Dark matter annihilations}
\label{sec:annihilations}

\subsection{Boltzmann equation}

The dark matter temperature evolution receives contributions from both the inelastic conversion $\chi L\leftrightarrow\bar\chi\Psi$ and the elastic scattering $\chi\phi\to\chi\phi$. We include both collision terms and focus on the parameter region where inelastic conversion dominates momentum exchange. The distribution function $f_\chi(t,p)$ obeys the Boltzmann equation in an expanding FRW background~\cite{Kolb:1990vq,Gondolo:1990dk,Binder:2017rgn,Bringmann:2006mu},
\begin{equation}
E\left(\frac{\partial f_\chi}{\partial t}-H \vec p\cdot \frac{\partial f_\chi}{\partial \vec p}\right)
=
C_{\rm ann}[f_\chi]+C_{\rm inel}[f_\chi]+C_{\chi\phi}[f_\chi].
\label{eq:Boltzmann_full}
\end{equation}
Here, the annihilation term $C_{\rm ann}[f_\chi]$ accounts for the $\chi\chi\to L\bar\Psi$ process, $C_{\rm inel}[f_\chi]$ describes the inelastic conversion scattering $\chi L\leftrightarrow\bar\chi\Psi$, and $C_{\chi\phi}[f_\chi]$ describes elastic scattering on the equilibrium mediator population. The elastic term conserves dark matter number and contributes to the temperature equation through its second moment. In general, a Fokker--Planck description is justified only when the typical momentum transfer per collision is small compared to the characteristic dark matter momentum, $q_{\rm tr}\ll p_\chi\sim \sqrt{m_\chi T_\chi}$, which is naturally realized for scatterings with a much lighter bath particle and sufficiently separated scales~\cite{Bringmann:2006mu,Binder:2017rgn,Binder:2021bmg}.
Although the incoming bath lepton $L$ is light, the inelastic conversion $\chi L\leftrightarrow\bar\chi\Psi$ involves a finite mass gap $\Delta m\equiv m_\Psi-m_L$. The production of the heavier final state can entail momentum transfers that are not small compared with the characteristic dark matter momentum, so the small momentum transfer assumption cannot be justified solely by the light incoming bath particle.
The Fokker--Planck expansion is therefore not reliable in the parameter region of interest, so we use the full inelastic collision operator. We assume a charge-symmetric dark matter population, $f_\chi(p)=f_{\bar\chi}(p)$, and evolve its number density and kinetic temperature.
\begin{equation}
C_{\rm inel}
=
E\int \frac{d^3\tilde p}{(2\pi)^3}\,
W_{\rm inel}(E,\tilde E,T)
\left[
e^{-(E-\tilde E)/(2T)} f_\chi(\tilde E)
-
e^{(E-\tilde E)/(2T)} f_\chi(E)
\right].
\label{eq:Cinel_full}
\end{equation}
The kernel $W_{\rm inel}$ is defined by
\begin{equation}
W_{\rm inel}
\equiv
\frac{e^{-(E-\tilde E)/(2T)}}{4 g_\chi E \tilde E}
\int \frac{d^3k}{(2\pi)^3\,2\omega}
\int \frac{d^3\tilde k}{(2\pi)^3\,2\tilde\omega}
\,f_L^{\rm eq}(\omega)\,
\left[1-f_\Psi^{\rm eq}(\tilde\omega)\right]
(2\pi)^4
\delta^{(4)}(p+k-\tilde p-\tilde k)\,
\overline{\left|\mathcal M_{\chi L\to\bar\chi\Psi}\right|^2},
\label{eq:Winel_full}
\end{equation}
where
\begin{equation}
\omega=\sqrt{k^2+m_L^2},
\qquad
\tilde\omega=\sqrt{\tilde k^2+m_\Psi^2},
\qquad
f_i^{\rm eq}(E_i)=\frac{1}{e^{E_i/T}+1}.
\label{eq:thermal_dist_inel}
\end{equation}

Since both $L$ and $\Psi$ are kept in thermal equilibrium with the plasma, the collision term obeys detailed balance. In particular, for a thermal dark matter distribution $f_\chi\propto e^{-E/T}$, the forward and inverse conversion rates cancel each other and one obtains $C_{\rm inel}=0$.
Integrating Eq.~\eqref{eq:Boltzmann_full} with $g_\chi\int d^3p/(2\pi)^3E^{-1}$ and
$g_\chi\int d^3p/(2\pi)^3\,p^2E^{-2}/3$ gives the zeroth and second moment equations.

For a fixed representative initial state $\chi\chi$, let $\gamma_{\chi\chi\to L_a\bar\Psi_a}^{(1)}$ and $\gamma_{\chi\chi\to\bar L_a\Psi_a}^{(1)}$ denote one fixed weak component and one fixed final charge channel. The complete annihilation reaction density sums both weak components and both final charge channels,
\begin{align}
\gamma_{\rm ann}
\equiv\sum_{a=1}^{2}\left[
\gamma_{\chi\chi\to L_a\bar\Psi_a}^{(1)}
+\gamma_{\chi\chi\to\bar L_a\Psi_a}^{(1)}
\right]
\simeq4\gamma_{\rm ann}^{(1)}.
\label{eq:gamma_ann_single_charge}
\end{align}
The last equality applies to the CP-even part of the collision term. Here $x\equiv m_\chi/T$, $Y_i\equiv n_i/s$, and $s=(2\pi^2/45)g_s^{\mathrm{eff}}(T)T^3$ is the entropy density, with $g_s$ in the equilibrium yields below denoting $g_s^{\mathrm{eff}}(T)$. The rescaled expansion rate is $\widetilde H\equiv H\big/[1+\frac13\,\mathrm{d}\ln g_s^{\mathrm{eff}}/\mathrm{d}\ln T\big]$.
\begin{align}
\label{BE}
\frac{dY_\chi}{dx}={}&
-\frac{1}{x s\widetilde H(T)}
\left[
\frac{Y_\chi^2}
{\left[Y_\chi^{\rm eq}(T_\chi)\right]^2}
\gamma_{\rm ann}(T_\chi)
-\gamma_{\rm ann}(T)
\right]
+\frac{12\zeta(3)}{\pi^2}
\frac{\epsilon_{\rm CP}\,\xi(T)Y_{\Delta L_a}}
{Y_\gamma\,x s\widetilde H(T)}
\gamma_{\rm ann}(T),\\
\frac{dY_{\Delta L_a}}{dx}={}&
\frac{\eta(T)}{x\widetilde H(T)}
\left\{
\frac{\epsilon_{\rm CP}}{2s}
\left[
\frac{Y_\chi^2}
{\left[Y_\chi^{\rm eq}(T_\chi)\right]^2}
\gamma_{\rm ann}(T_\chi)
-\gamma_{\rm ann}(T)
\right]
-\frac{24\zeta(3)}{\pi^2}\xi(T)
\mathcal W_{\Delta L_a}(x)Y_{\Delta L_a}
\right\}
\nonumber\\
&{}+\frac{d\ln\eta(T)}{dx}Y_{\Delta L_a}.
\label{eq:BE_asymmetry}
\end{align}
Here $\xi(T)$ controls the chemical-potential combination entering the washout term, while $\eta(T)$ accounts for spectator redistribution and relates the lepton-doublet asymmetry to the Standard Model $B-L$ asymmetry through $Y_{\Delta L}=-\eta(T)Y_{B-L}^{\rm SM}$. Both factors are defined in Eq.~\eqref{eq:visible_bl}.
The quantity $Y_\chi^{\mathrm{eq}}(T_\chi)$ is the equilibrium yield for one charge, evaluated at the dark matter temperature, and is used only to normalize the nonequilibrium distribution. We close the moment hierarchy with the Maxwell--Boltzmann ansatz $f_\chi(p)\simeq[n_\chi/n_\chi^{\mathrm{eq}}(T_\chi)]e^{-E/T_\chi}$, normalized to the actual one-charge density. The present calculation therefore retains the full inelastic collision kernel within a coupled two-moment treatment, rather than solving the full phase-space evolution of $f_\chi$. Resolving possible non-Maxwellian distortions near resonance requires a full phase-space treatment, which we leave for future work.
\begin{figure}[htbp]
	\begin{center}
		 
		\includegraphics[width=8cm,height=4cm]{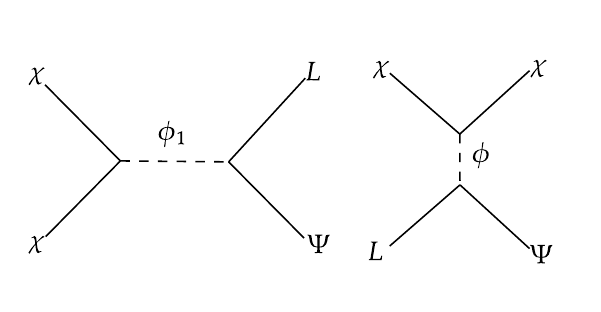}
	\end{center}
	\caption{
Feynman diagrams for dark matter annihilation and scattering processes: (left) $\chi\chi \to L\bar\Psi$ via an $s$ channel mediator $\phi$, (right) $t$ channel inelastic scattering $\chi L \to \bar\chi \Psi$ for kinetic decoupling.
  }
	\vspace{-0.5cm}
	\label{fig:cBE}
\end{figure}

The number density Boltzmann equation (nBE) treatment is recovered in the kinetic equilibrium limit $T_\chi=T$, where the separate evolution of $y$ is not included. The process $\chi\chi\to L\bar\Psi$ enters both the abundance equation and the temperature equation for $y$. The latter involves the thermal average weighted by the second moment, $\langle\sigma v\rangle_{2,T}$, whose explicit form is given in the \hyperref[app:thermal_avg]{Appendix}. Here $\langle\sigma v\rangle_2$ denotes one fixed weak component and one fixed final charge channel.

In this framework, the dominant annihilation channels are $\chi\chi\to L\bar\Psi$. The annihilation and conversion processes are illustrated in Fig.~\ref{fig:cBE}. The quantity $T_\chi$ is defined from the second moment of the dark matter distribution~\cite{Gondolo:2012vh, Visinelli:2015eka},
\begin{align}
T_\chi \equiv \frac{g_\chi}{n_\chi}\int \frac{d^3p}{(2\pi)^3}\frac{p^2}{3E}\,f_\chi(p),
\qquad
y \equiv \frac{m_\chi T_\chi}{s^{2/3}},
\qquad
y_{\rm eq} \equiv \frac{m_\chi T}{s^{2/3}}.
\end{align}

The inelastic conversion contribution to the second moment is denoted by $\mathcal{F}_{\rm conv}$~\cite{Binder:2017rgn}, while the elastic contribution from $\chi\phi\to\chi\phi$ is denoted by $\mathcal{F}_{\chi\phi}$. Let $\mathcal{F}_{\rm conv}^{(1)}$ denote one fixed weak component and one bath charge channel, for example $\chi L_a\leftrightarrow\bar\chi\Psi_a$. The Hermitian conjugate Yukawa interaction also gives $\chi\bar L_a\leftrightarrow\bar\chi\bar\Psi_a$, which acts on the same representative $f_\chi$ distribution. In a charge-symmetric bath, the two bath charge channels have equal kernels, so
\begin{align}
\mathcal{F}_{\rm conv}
=4\mathcal{F}_{\rm conv}^{(1)}.
\end{align}
We denote the corresponding elastic contribution by $\mathcal{F}_{\chi\phi}(x,y)\equiv\left.dy/dx\right|_{\chi\phi}$, evaluated with the equilibrium mediator distribution and the full elastic collision integral. The temperature equation then reads
\begin{align}
\frac{y'}{y}
=&
\frac{\mathcal{F}_{\rm conv}(x,y)+\mathcal{F}_{\chi\phi}(x,y)}{y}
+\frac{4sY_\chi}{x\widetilde H}
\Bigl[
\frac{[Y_\chi^{\rm eq}(T)]^2}{Y_\chi^2}\frac{y_{\rm eq}}{y}\langle\sigma v\rangle_{2,T}
-\langle\sigma v\rangle_{2,T_\chi}
\Bigr]
-\frac{Y_\chi'}{Y_\chi}
+\frac{H}{x\widetilde H}\frac{\langle p^4/E^3\rangle}{3T_\chi} 
\label{eq:y_fullscatt}
\end{align}
Linearizing the inelastic conversion contribution around kinetic equilibrium gives the corresponding effective momentum transfer rate. Defining
\begin{align}
\Delta y \equiv y-y_{\rm eq},
\qquad
\left.\frac{dy}{dx}\right|_{\rm conv}\equiv \mathcal{F}_{\rm conv}(x,y),
\end{align}
one finds near equilibrium
\begin{align}
\left.\frac{d\Delta y}{dx}\right|_{\rm conv}
\simeq
\left.
\frac{\partial \mathcal{F}_{\rm conv}(x,y)}{\partial y}
\right|_{y=y_{\rm eq}}
\Delta y.
\label{eq:linear_scatt}
\end{align}
Comparing this with the standard relaxation form
\begin{align}
\left.\frac{d\Delta y}{dt}\right|_{\rm conv}
=
    -2\gamma_{\rm conv}(T)\,\Delta y,
\qquad
\frac{dx}{dt}\simeq \tilde H x,
\end{align}
For one fixed weak component and one bath charge channel, we define the corresponding effective momentum transfer rate by
\begin{align}
    \frac{\gamma^{\rm eff}(x)}{\tilde H(x)}
\equiv
-\frac{x}{2}
\left.
\frac{\partial \mathcal{F}_{\rm conv}^{(1)}(x,y)}{\partial y}
\right|_{y=y_{\rm eq}}.
\label{eq:gammaMT_eff_def}
\end{align}
The same linearization procedure is used to obtain $\gamma_{\chi\phi}$ from the full elastic collision integral.
For a charge-symmetric bath, summing over the weak components and bath charge channels gives $\gamma_{\rm conv}=4\gamma^{\rm eff}$. Including the elastic contribution, the total momentum relaxation rate is $\gamma_{\rm kin}=\gamma_{\rm conv}+\gamma_{\chi\phi}$. Although both scattering processes are included in the second moment equation, inelastic conversion provides the dominant contribution to momentum relaxation near the reference decoupling point in the selected parameter region, where $\gamma_{\chi\phi}/\gamma_{\rm conv}<0.3$ as specified in Sec.~\ref{sec:pheno}. Thus, $\gamma_{\rm kin}\simeq\gamma_{\rm conv}$ in this regime. The kinetic decoupling temperature is determined using the total rate,
\begin{align}
\gamma_{\rm kin}(x_{\rm kd})
=\gamma_{\rm conv}(x_{\rm kd})+\gamma_{\chi\phi}(x_{\rm kd})
\simeq\tilde H(x_{\rm kd}).
\label{eq:xkd_def_new}
\end{align}
The equilibrium abundances of $\chi$, $L$, and $\Psi$ in the Maxwell--Boltzmann approximation, together with the photon yield, are
\begin{equation}
Y_\chi^{\mathrm{eq}} = \frac{45 x^2}{2\pi^4 g_s}\,\mathcal{K}_2(x), \qquad
Y_i^{\mathrm{eq}} = \frac{45 g_i x^2}{4\pi^4 g_s}
\left(\frac{m_i}{m_\chi}\right)^2
\mathcal{K}_2\!\left( \frac{m_i}{m_\chi} x \right),\qquad
Y_\gamma = \frac{45\,\zeta(3)}{\pi^4 g_s}\,.
\end{equation}
Here $K_2$ denotes the modified Bessel function of the second kind. The thermally averaged interaction rates for decay and annihilation processes read \cite{Plumacher:1996kc}
\begin{align}
\gamma_{1\to 2+3} &= n_1^{\rm eq}\,\frac{K_1}{K_2}\,\Gamma_{1\to 2+3}\,,\\
\gamma_{1+2\to 3+4} &= \frac{T}{64\pi^4}\int_{(m_1+m_2)^2}^{\infty} \mathrm{d}s'\,
\hat\sigma_{1+2\to 3+4}\,\sqrt{s'}\,K_1\!\left(\frac{\sqrt{s'}}{T}\right)\,.
\end{align}
The reduced cross section is defined as
\[
\widehat{\sigma}_{1+2\to 3+4}(s')
=\frac{2\lambda(s',m_1^2,m_2^2)}{s'}\,
\sigma_{1+2\to 3+4}(s') .
\]

\subsection{Leptogenesis}

\begin{figure}[H]
    \centering
    \includegraphics[width=0.90\textwidth]{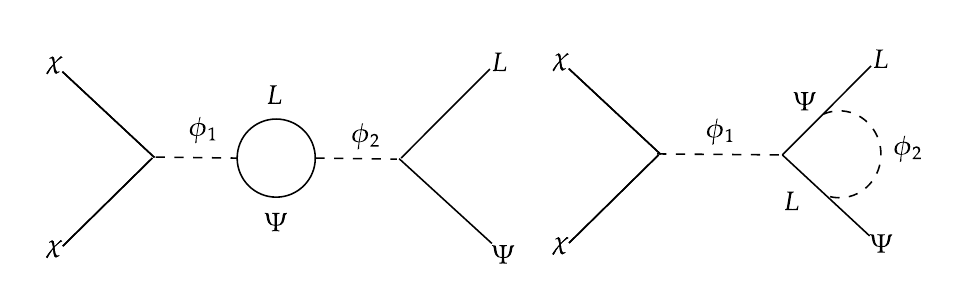}
    \caption{Representative one loop contributions to $\chi\chi\to L\bar\Psi$: the pseudoscalar self-energy correction (left) and the vertex correction (right). Their interference with the tree level amplitude generates the CP asymmetry.}
    \label{fig:cp_loops}
\end{figure}

In this model, a lepton asymmetry can arise through the annihilation process $\chi\chi\to L\bar\Psi$. The CP violating phases arise from the interference between the tree level amplitude and the one loop self energy and vertex corrections involving the heavier mediator \( \phi_2 \), as illustrated in Fig.~\ref{fig:cp_loops}.

The spectator factors and visible sector asymmetry relations are given below~\cite{Harvey:1990qw,Nardi:2006fx}.
This conversion is valid only while electroweak sphalerons remain active~\cite{Kuzmin:1985mm}. We denote the sphaleron freeze-out temperature by $T_\ast\simeq 131.7~{\rm GeV}$~\cite{DOnofrio:2014rug} and define
\[
x_\ast\equiv \frac{m_\chi}{T_\ast}.
\]
We use $x_{\rm lep}$ to characterize the end of the main asymmetry-production stage. We impose the conservative requirement $x_{\rm lep}<x_\ast$ to ensure that asymmetry generation is completed before sphaleron freeze-out.
Spectator effects redistribute the asymmetry between the SM lepton doublet $L$ and the vectorlike doublet $\Psi$~\cite{Cui:2011ab}. We evolve the asymmetry in one weak component, $Y_{\Delta L_a}\equiv(n_{L_a}-n_{\bar L_a})/s$. The spectator coefficients and asymmetry relations are
{\small
\begin{equation}
\begin{gathered}
(\xi(T),\eta(T))=
\begin{cases}
\displaystyle \left(1,\frac{442}{711}\right), & {\rm post},\\[4pt]
\displaystyle \left(\frac{968+223\kappa_\Psi}{221+223\kappa_\Psi},
\frac{8(221+223\kappa_\Psi)}{9(316+329\kappa_\Psi)}\right), & {\rm eq},
\end{cases}\\
Y_{\Delta L}=2Y_{\Delta L_a}=-\eta(T)Y_{B-L}^{\rm SM},\qquad
Y_{\Delta B}=c_B(T)Y_{B-L}^{\rm SM},\\
c_B(T)=
\begin{cases}
\displaystyle \frac{28}{79}, & {\rm post},\\[4pt]
\displaystyle \frac{8(14+13\kappa_\Psi)}{316+329\kappa_\Psi}, & {\rm eq},
\end{cases}
\qquad
Y_{\Delta L}^{\rm SM}=[c_B(T)-1]Y_{B-L}^{\rm SM}.
\end{gathered}
\label{eq:visible_bl}
\end{equation}
}
The susceptibility, chemical-potential constraints, and reconstructed sector asymmetries are given in the \hyperref[app:spectator]{Appendix}. The two limits are selected according to the $n_R$ decoupling time, with an instantaneous switch at $x_{n,\rm dec}$ when decoupling occurs during asymmetry production.
In addition, the washout rate is dominated by the leading lepton number violating processes and can be expressed as
\begin{align}
    \mathcal W_{\Delta L_a}(x)
    \equiv \Gamma_{\rm washout}(x)=
\frac{\frac14\gamma_{\rm ann}+\frac12\gamma_{L\Psi}+\frac12\gamma_{\Psi\Psi}}{n_{\gamma}}
+\frac{n_\chi}{n_\chi^{\mathrm{eq}}}
\frac{\gamma^{(1)}_{\chi L\to\bar\chi\Psi}(T_\chi,T)}{n_\gamma}
    \label{Eq:washout}
\end{align}
Here $\gamma^{(1)}_{\chi L\to\bar\chi\Psi}(T_\chi,T)$ denotes the conversion reaction density for one weak component and one bath charge channel.

In the parameter region considered here, washout is dominated by the channel $L\bar\Psi \to \bar L \Psi$, which violates lepton number. Summing over the initial states $\chi\chi$ and $\bar\chi\bar\chi$ and both final charge channels for one component of the weak doublet, we define the CP asymmetry as
\begin{align}
\epsilon_{\mathrm{CP}} =
\frac{
\sigma(\chi\chi\to L\bar\Psi)+\sigma(\bar\chi\bar\chi\to L\bar\Psi)
-\sigma(\chi\chi\to\bar L\Psi)-\sigma(\bar\chi\bar\chi\to\bar L\Psi)}
{\sigma(\chi\chi\to L\bar\Psi)+\sigma(\bar\chi\bar\chi\to L\bar\Psi)
+\sigma(\chi\chi\to\bar L\Psi)+\sigma(\bar\chi\bar\chi\to\bar L\Psi)} .
\end{align}
In the present work, $\epsilon_{\mathrm{CP}}$ is treated as an effective phenomenological parameter entering the Boltzmann equations. As a reference estimate, the one-loop calculation in the more general coupling framework of Ref.~\cite{Cui:2011ab}, under the hierarchical mediator approximation and for large CP-violating phases, gives
\begin{equation}
\epsilon_{\max}
\equiv
\frac{2\lambda_L^2}{3\pi\sqrt{5}}
\frac{m_\chi^2}{m_\phi^2}
\left[
7-15\left(\frac{m_\Psi}{2m_\chi}\right)^2
+9\left(\frac{m_\Psi}{2m_\chi}\right)^4
-\left(\frac{m_\Psi}{2m_\chi}\right)^6
\right].
\label{eq:epsilon-max}
\end{equation}
Equation~\eqref{eq:epsilon-max} is used as a reference estimate, rather than a one-loop result derived for the restricted pseudoscalar coupling structure considered here.

\subsection{Benchmark Points}

\begin{figure}[htbp]
	\begin{center}
		 
		\includegraphics[width=6.5cm,height=6.5cm]{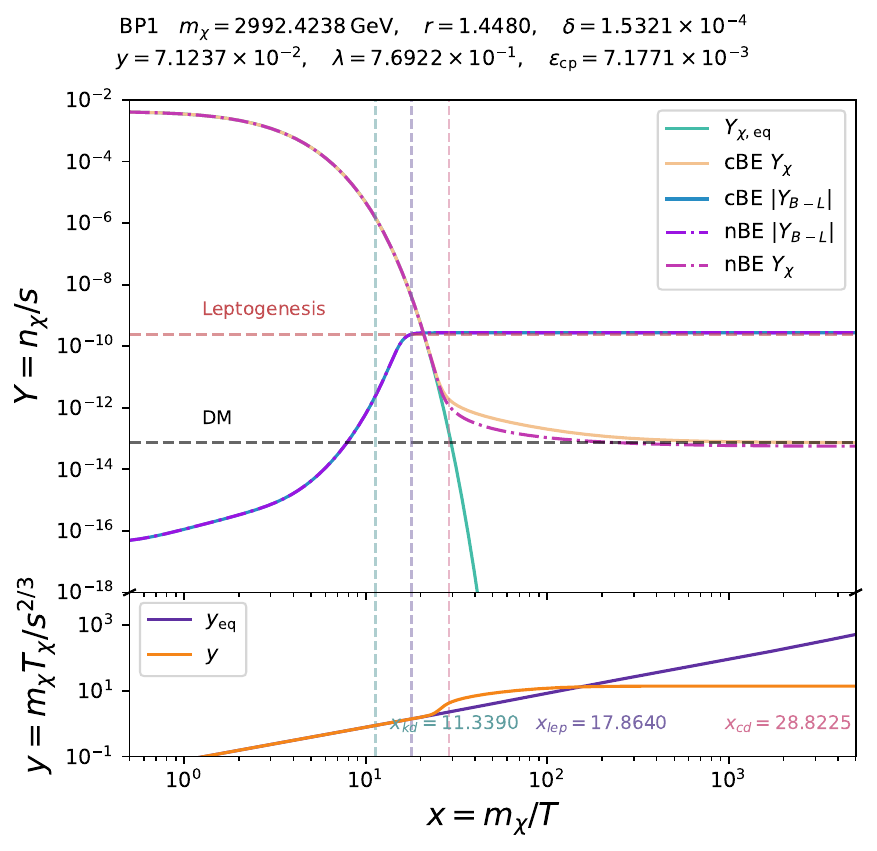}
        \includegraphics[width=6.5cm,height=6.5cm]{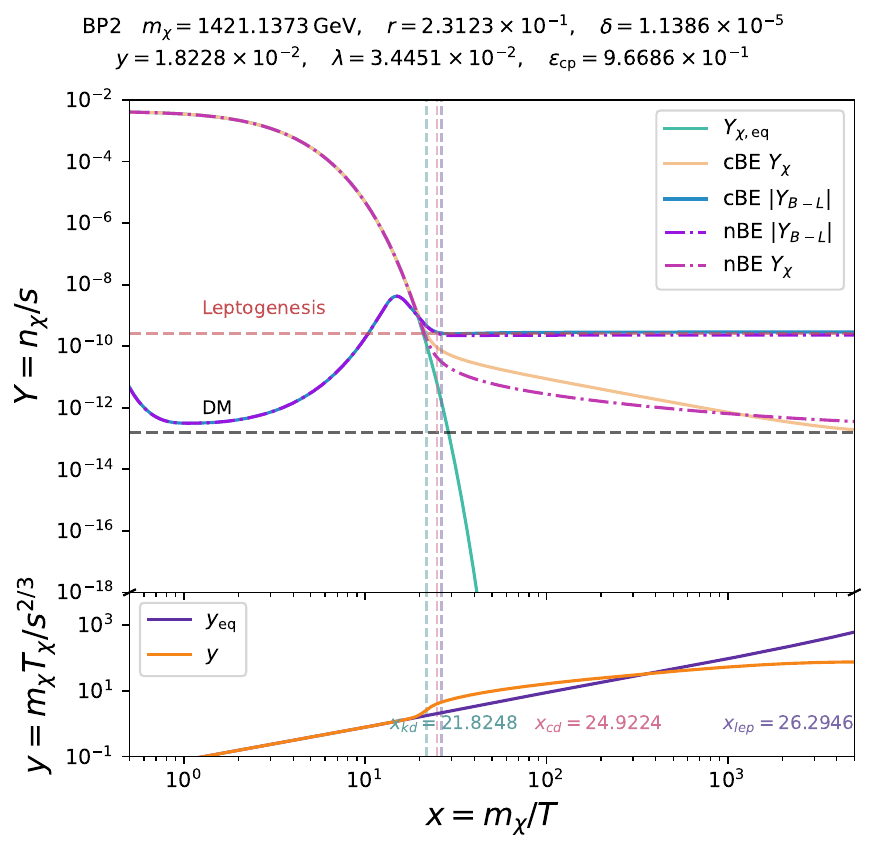}

        \includegraphics[width=6.5cm,height=6.5cm]{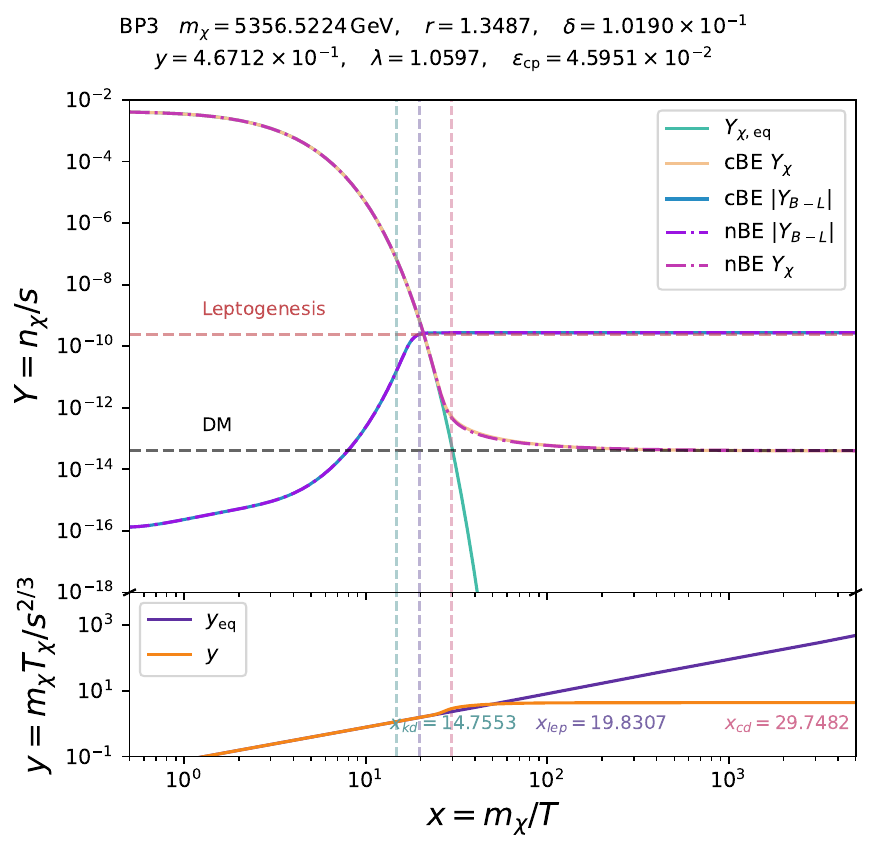}
	\end{center}
	\caption{
Evolution of $Y_\chi$, $|Y_{B-L}|$, and $y=m_\chi T_\chi s^{-2/3}$ for the benchmark points BP1--BP3. 
The cBE and nBE results are compared in the upper panels, and the temperature variable is shown in the lower panels. 
Vertical dashed lines mark $x_{\rm kd}$, $x_{\rm lep}$, and $x_{\rm cd}$. 
The horizontal lines show the $B-L$ yield corresponding to the observed baryon asymmetry and the dark matter yield corresponding to $\Omega_{\rm DM}h^2\simeq 0.12$. 
If $x_{\rm lep}>x_\ast$, the plotted asymmetry represents late time $B-L$ evolution rather than a sphaleron converted baryon asymmetry.
}
	\vspace{-0.5cm}
	\label{fig:YY}
\end{figure}

We consider the three benchmark points shown in Fig.~\ref{fig:YY} and parametrize the mediator mass as $m_\phi=2m_\chi/\sqrt{1+\delta}$, where $\delta=4m_\chi^2/m_\phi^2-1$ measures the distance from resonance. Kinetic equilibrium is maintained mainly by the inelastic conversion $\chi L\leftrightarrow\bar\chi\Psi$, whose rate becomes increasingly Boltzmann suppressed as the thermal population loses sufficient energy to produce the heavier $\Psi$ state. The resulting kinetic decoupling time is controlled by the mass ratio $r=m_\Psi/m_\chi$, the conversion couplings, and the resonance condition. All three benchmarks considered here kinetically decouple before chemical decoupling, but their relic abundance and asymmetry responses differ. Near resonance, the observed relic abundance can be obtained with smaller couplings, which further suppresses the conversion rate and can move $x_{\rm kd}$ to much earlier times. Early kinetic decoupling alone, however, does not guarantee a sizable change in the relic abundance. A visible cBE correction requires the annihilation rate to remain sensitive to the dark matter momentum distribution after kinetic decoupling.
We define a reference chemical decoupling point using the annihilation rate evaluated on the thermal-bath equilibrium background,
\[
\Gamma_{\rm cd}(x)\equiv
\frac{\gamma_{\rm ann}(x)}
{n_\chi^{\rm eq}(x)}
=n_\chi^{\rm eq}(x)
\left\langle\sigma v\right\rangle_{\rm ann}(x).
\]
We determine the reference chemical decoupling point from $\Gamma_{\rm cd}(x_{\mathrm{cd}})\simeq H(x_{\mathrm{cd}})$, and the kinetic decoupling point from $\gamma_{\rm kin}(x_{\mathrm{kd}})\simeq\widetilde H(x_{\mathrm{kd}})$~\cite{Gondolo:1990dk,Kolb:1990vq,Berezinsky:2003vn,Chen:2001jz,Duch:2017nbe}. The washout process freezes out when the damping rate including spectator factors, $[24\zeta(3)/\pi^2]\eta\xi\Gamma_{\rm washout}$, drops below the Hubble rate. In the strong washout regime, $x_{\mathrm{w}}$ is usually close to the time when the final asymmetry is effectively fixed. For weaker washout, the end of leptogenesis is instead controlled by the combined evolution of the source and washout terms and need not coincide with $x_{\mathrm{w}}$. We use $x_{\rm lep}$ to characterize the end of the main asymmetry-production stage. In the numerical analysis we solve the asymmetry equation, Eq.~\eqref{eq:BE_asymmetry}, with the initial condition $Y_{\Delta L_a}(x\ll 1)=0$, corresponding to a vanishing primordial lepton asymmetry.
The benchmarks illustrate how early kinetic decoupling modifies the relic abundance and the late time evolution of the $B-L$ asymmetry. When leptogenesis occurs relatively late, the main correction is to the late evolution of $B-L$. The final baryon asymmetry changes only if sphaleron conversion remains effective during this period.

For BP1, the small positive resonance parameter $\delta=1.5321\times10^{-4}$ places the mediator pole just below the $\chi\chi$ threshold and makes the annihilation rate strongly velocity dependent. Kinetic decoupling occurs at $x_{\rm kd}\simeq11.3390$, earlier than both $x_{\rm lep}\simeq17.8640$ and chemical decoupling at $x_{\rm cd}\simeq28.8225$. After kinetic decoupling, conversion can no longer efficiently restore kinetic equilibrium, while annihilation remains active. Since $s\simeq4m_\chi^2+m_\chi^2v_{\rm rel}^2$ and $m_\phi<2m_\chi$, dark matter pairs with smaller relative velocities lie closer to the Breit--Wigner pole and therefore have larger annihilation rates. The annihilation kernel consequently removes less kinetic energy relative to particle number than a depletion independent of momentum would. The surviving distribution develops a larger second moment, which appears as a rise of $y$ after $x_{\rm kd}$. The temperature enhancement persists into the residual annihilation stage, after which $y$ tends toward the nearly constant behavior expected for a decoupled nonrelativistic species. The relic abundance ratio is $\Omega_\chi^{\rm cBE}h^2/\Omega_\chi^{\rm nBE}h^2\simeq1.3028$. The lepton asymmetry is much less affected, with the cBE/nBE ratio remaining close to unity, because the largest temperature and abundance differences develop after the main asymmetry production stage.

For BP2, the very small value $\delta=1.1386\times10^{-5}$ places the annihilation process extremely close to resonance, while the smaller mass ratio $r\simeq0.231$ makes the conversion less endothermic than in BP1 and BP3. The characteristic times follow the ordering $x_{\rm kd}\simeq21.8248<x_{\rm cd}\simeq24.9224<x_{\rm lep}\simeq26.2946$. During the late asymmetry evolution, the source depends on both the temperature dependent annihilation rate and the dark matter abundance through
\begin{equation}
\frac{Y_\chi^2}{[Y_\chi^{\rm eq}(T_\chi)]^2}\,\gamma_{\rm ann}(T_\chi) - \gamma_{\rm ann}(T),
\end{equation}
The asymmetry ratio is $|Y_{B-L}^{\rm cBE}(x_{\rm lep})|/|Y_{B-L}^{\rm nBE}(x_{\rm lep})|\simeq1.0992$ and reaches $|Y_{B-L}^{\rm cBE}|/|Y_{B-L}^{\rm nBE}|\simeq1.2576$ at late times. The relic abundance ratio is $\Omega_\chi^{\rm cBE}h^2/\Omega_\chi^{\rm nBE}h^2\simeq4.9117\times10^{-1}$. BP2 does not satisfy the sphaleron condition $x_{\rm lep}<x_\ast$, but is retained to illustrate late kinetic feedback.

For BP3, inelastic conversion drives kinetic decoupling before chemical decoupling, with $\delta=1.0190\times10^{-1}$ and $x_{\rm kd}\simeq14.7553<x_{\rm lep}\simeq19.8307<x_{\rm cd}\simeq29.7482$. This point lies farther from the narrow resonant region than BP1 and BP2. Its temperature evolution shows a more modest departure from equilibrium, and $y$ approaches the nearly constant behavior expected for a freely cooling nonrelativistic species at late times. Despite early kinetic decoupling, the cBE and nBE relic abundances remain close, with $\Omega_\chi^{\rm cBE}h^2/\Omega_\chi^{\rm nBE}h^2\simeq9.8226\times10^{-1}$. The lepton asymmetry is likewise almost unchanged, with its cBE/nBE ratio remaining close to unity; the largest temperature departures develop after the main asymmetry production stage.

The mass ratio $r \equiv m_\Psi/m_\chi$, the resonance parameter $\delta$, and the dark matter mass $m_\chi$ control the relative locations of chemical decoupling, kinetic decoupling, and the effective end of leptogenesis. Varying these parameters changes how strongly the annihilation, source, and washout terms sample the nonthermal stage. The relic abundance is most sensitive when kinetic decoupling occurs before chemical decoupling and the annihilation cross section retains a strong temperature dependence, as in the resonant regime. By contrast, the lepton asymmetry is more sensitive when its production overlaps with a substantial departure from kinetic equilibrium, particularly if it continues through chemical decoupling. If $x_{\mathrm{lep}} < x_{\mathrm{cd}}$, the asymmetry is already close to its final value before chemical decoupling, so dark matter temperature corrections induce only a mild net modification in $Y_{B-L}$.

\section{Parameter Space and Results}
\label{sec:pheno}

Throughout the scan we impose $m_\chi < m_\phi < 2m_\chi$ and $m_\Psi + m_L < 2m_\chi$. Using $\delta \equiv 4m_\chi^2/m_\phi^2 - 1$, the first condition is equivalent to $0 < \delta < 3$. This choice forbids the tree-level decay $\phi \to \chi\chi$ and closes the annihilation production channel $\chi\bar\chi\to\phi\phi$ at threshold, although a thermal tail can still open this forbidden-annihilation channel. The condition $m_\Psi+m_L<2m_\chi$ keeps the annihilation channel $\chi\chi\to L\bar\Psi$ kinematically open. We additionally require $m_\Psi<m_\phi$ to kinematically close the competing decay $\Psi\to\phi L$ and ensure transfer of the $\Psi$ asymmetry through $\Psi\to Hn_R$. To identify the viable parameter space, we independently scan
\begin{align}
m_\chi \in [200,10^4]~\mathrm{GeV}, \qquad
r \equiv \frac{m_\Psi}{m_\chi} \in [0.1,2], \qquad
\delta \in [10^{-5},0.5), \qquad
y_\chi,\ \lambda_L \in [10^{-4},3] .
\end{align}

We require $\gamma_{\chi\phi}/\gamma_{\rm conv}<0.3$ at the reference kinetic decoupling point, defined by the total momentum relaxation rate crossing $\gamma_{\rm kin}=\widetilde H$. We also require the estimated mediator-decay contribution to the lepton asymmetry at sphaleron freeze-out, $x_\ast$, to remain below $1\%$ of the asymmetry required by observation, using a linear response estimate with $|\epsilon_\phi|\leq1$. These conditions select the parameter region where inelastic conversion dominates momentum exchange near the reference decoupling point and the independent mediator-decay CP source can be neglected within this estimate. Including elastic scattering in the coupled Boltzmann equations shifts the kinetic decoupling temperature but has little effect on the final relic abundance and lepton asymmetry in the benchmark comparisons.

We fix $\lambda_n=10^{-6}$ and impose $m_\Psi>1.0~\mathrm{TeV}$ when identifying allowed points. The same coupling controls the decay $\Psi\to Hn_R$, the mixing between $\Psi$ and $n_R$ induced by electroweak symmetry breaking, the associated light conversion channels, decays of the charged states, and the dark radiation contribution. Throughout the relevant leptogenesis epoch, $\Psi\to Hn_R$ remains faster than the Hubble expansion, while $\gamma_{\rm light}/\widetilde H\ll1$ near kinetic decoupling. The light conversion channels therefore have a negligible effect on the dark matter thermal history. Since the relevant interactions conserve the global fermion charge, $Y_{\Delta n}(x_\ast)$ is fixed by the asymmetries in the visible and $\Psi$ sectors and is therefore determined from the relations given in the \hyperref[app:spectator]{Appendix} rather than evolved independently. The charged and neutral components decay through $\psi^\pm\to W^\pm n_R$ and $N\to Zn_R,hn_R$, with $c\tau_{\psi^\pm}<1~\mathrm{mm}$ and the $W^\pm n_R$ channel dominating over the pion and leptonic modes~\cite{Bhattacharya:2015qpa}. For $m_\Psi>1~\mathrm{TeV}$, the heavy doublet decays promptly, so disappearing-track searches targeting long-lived charged particles are not applicable~\cite{ATLAS:2026disappearing}.

We solve the coupled Boltzmann equations using a Python implementation based on the DRAKE framework~\cite{Binder:2021bmg} and scan the parameter space using Python for Monte Carlo sampling and the covariance matrix adaptation evolution strategy (CMA-ES). 
We retain parameter points with $\Omega_\chi^{\mathrm{cBE}} h^2 \in [0.11,0.13]$~\cite{Planck:2018vyg}, so that the relic density consistently includes the effect of early kinetic decoupling. Since the final asymmetry is approximately linear in the CP parameter, $Y_{B-L}\propto \epsilon_{\mathrm{CP}}$, we determine the value of $\epsilon_{\mathrm{CP}}$ required to reproduce the observed baryon asymmetry $Y_{\Delta B}^{\mathrm{obs}}$, rather than treating it as an independent scan parameter. We also require $|\epsilon_{\mathrm{CP}}|<1$. The allowed points must further satisfy $|\epsilon_{\mathrm{CP}}|\leq\epsilon_{\max}$, $x_{\rm lep}<x_\ast$, and $m_\Psi>1~\mathrm{TeV}$. In the scan figures, circles denote the allowed points, while stars denote relic density solutions that fail at least one of these additional conditions.

For the retained scan points, the equilibrium-parent estimate using Eqs.~\eqref{eq:nr-radiation-evolution}--\eqref{eq:nr-late-injection-neff} gives a total dark-radiation contribution, including the thermal component and the energy injected after $n_R$ decoupling, of $\Delta N_{\rm eff}^{\rm total}\in[5.9670\times10^{-2},\,7.1527\times10^{-2}]$. This remains below the current cosmological bound, $\Delta N_{\rm eff}<0.107$ at 95\% C.L.~\cite{Goldstein:2026}, and therefore does not further restrict the viable parameter space. Using Eq.~\eqref{eq:delta-h-expansion}, the corresponding expansion-rate correction is $\delta H/H\in[1.0225\times10^{-2},\,1.1572\times10^{-2}]$. This correction has little effect on the Boltzmann evolution in the parameter region considered.

\begin{figure}[htbp]
	\begin{center}
		 
		\includegraphics[width=8cm,height=6cm]{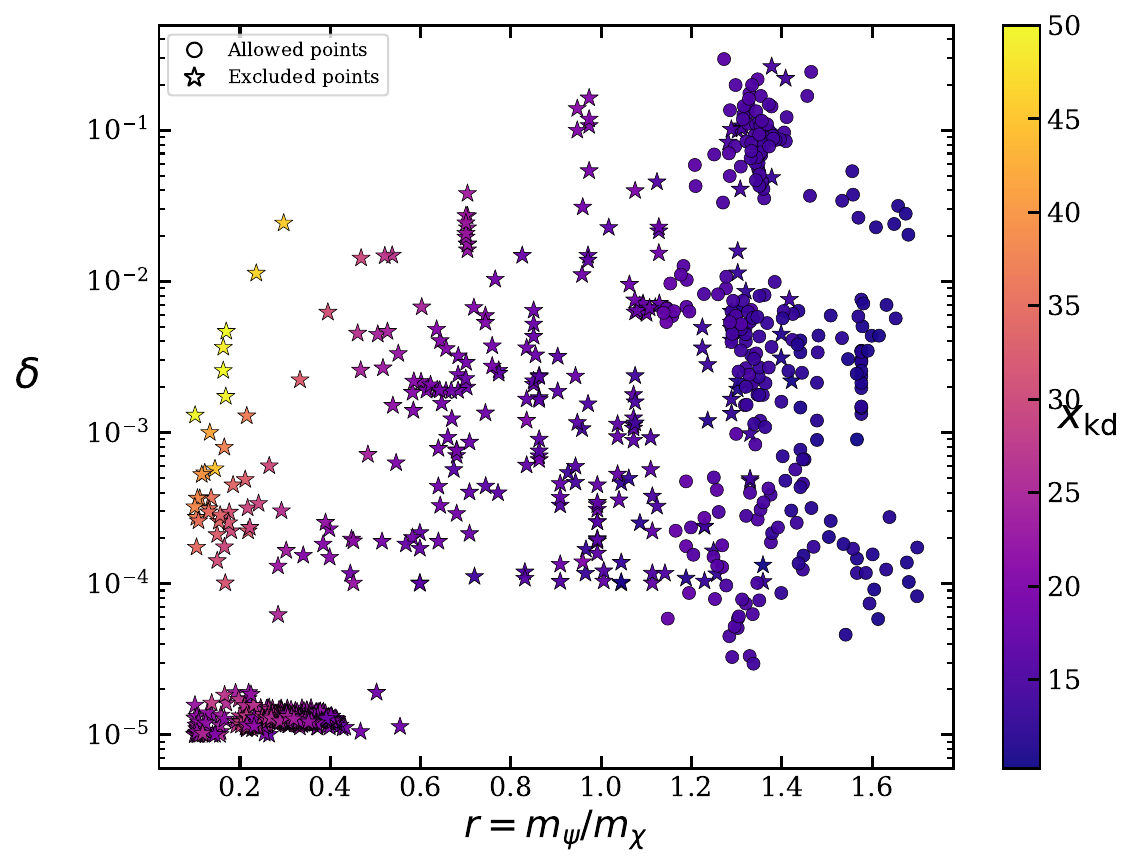}
        \includegraphics[width=8cm,height=6cm]{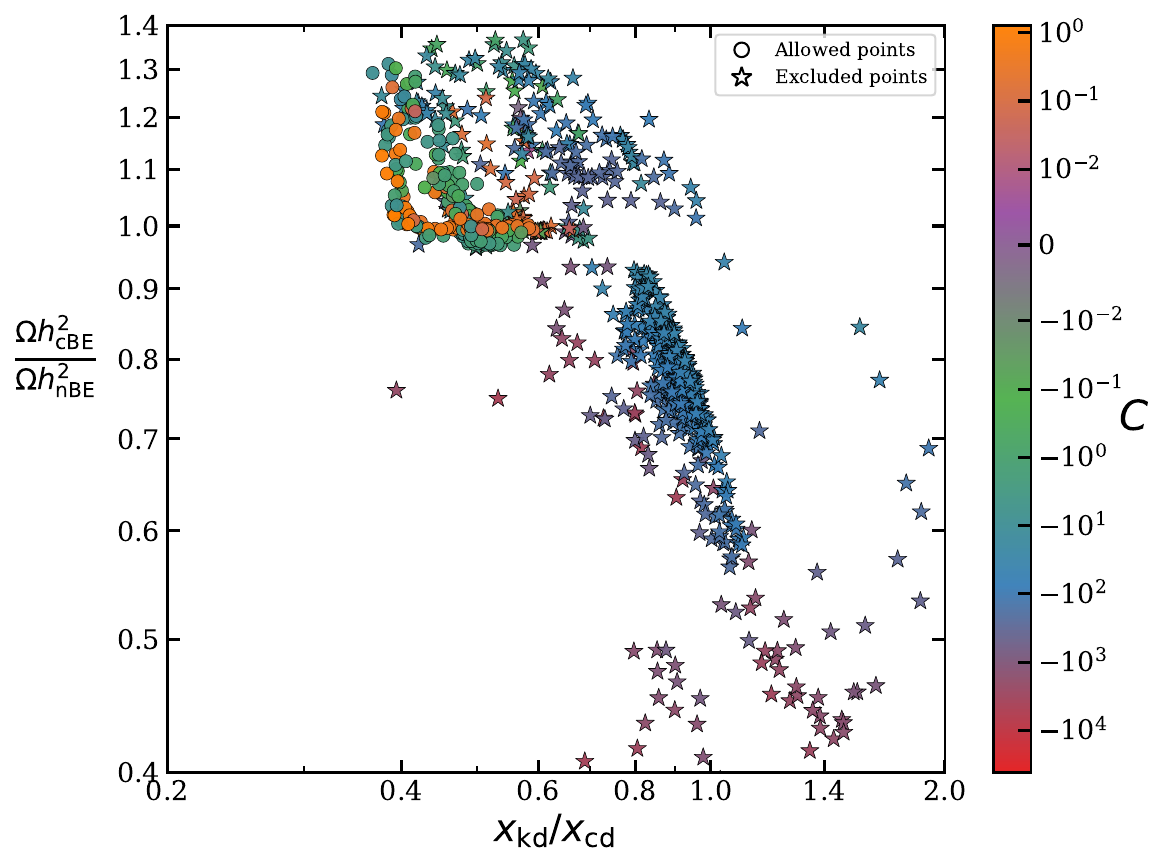}
	\end{center}
\caption{Left panel: scan points in the $r$--$\delta$ plane, colored by $x_{\rm kd}$. Right panel: ratio of relic densities, $\Omega_\chi^{\rm cBE}h^2/\Omega_\chi^{\rm nBE}h^2$, versus $x_{\rm kd}/x_{\rm cd}$, with the color scale showing the coefficient $C$ in $\sigma v\simeq\sigma_0(1+Cv^2)$.
}
	\vspace{-0.5cm}
	\label{fig:rdelta}
\end{figure}

Figure~\ref{fig:rdelta} shows how the kinetic history is controlled by the mass ratio $r=m_\Psi/m_\chi$ and the resonance parameter $\delta$. The points satisfying both the dark matter and leptogenesis requirements lie predominantly at $r>1$, where the conversion $\chi L\leftrightarrow\bar\chi\Psi$ involves a sizable mass gap and is strongly inelastic. The resulting Boltzmann suppression weakens momentum exchange and drives kinetic decoupling to $x_{\mathrm{kd}}\lesssim 25$, including points farther from the narrow resonant region. As $r$ decreases toward $0.1$, the conversion becomes less endothermic, momentum exchange remains efficient for longer, and kinetic decoupling is delayed to $x_{\mathrm{kd}}\gtrsim 40$. The distribution therefore reflects the characteristic role of inelastic conversion in producing early kinetic decoupling, while resonance effects can further shift $x_{\mathrm{kd}}$ through their impact on the couplings required by the relic density.

Near the annihilation resonance, $m_\phi\simeq2m_\chi$, the thermal abundance of the heavy mediator is strongly Boltzmann suppressed. Consequently, elastic scattering with the mediator can be less effective in maintaining kinetic equilibrium than inelastic conversion, despite the absence of an inelastic threshold. This provides a physical basis for conversion-dominated kinetic decoupling even when elastic scattering with thermal mediators is included.

The difference between the cBE and nBE results in the right panel of Fig.~\ref{fig:rdelta} can be traced to the momentum dependence of the annihilation cross section. In $s$-channel models, the Breit--Wigner propagator can induce a pronounced velocity dependence driven by resonance, which has been widely discussed in the context of early kinetic decoupling~\cite{Bi:2011qm,Binder:2021bmg}. In $\chi\chi\to L\bar\Psi$, the mediator is off shell because $m_\phi<2m_\chi$. For a single weak component and final charge channel, the laboratory-frame cross section is
\begin{equation}
\sigma v_{\rm lab}^{(1)} \;=\; 
\frac{y_\chi^{2}\lambda_{L}^{2}}{32\pi}\,
D(s)\,\alpha_\chi(s)\,
\frac{\sqrt{(s-(m_\Psi+m_{L})^{2})(s-(m_\Psi-m_{L})^{2})}}{s-2m_\chi^{2}},
\label{eq:sigma-ann}
\end{equation}
Here $\alpha_\chi(s)=s-m_\Psi^{2}-m_{L}^{2}$ and $D(s)=1/\big[(s-m_\phi^{2})^{2}+m_\phi^{2}\Gamma_\phi^{2}\big]$ denotes the Breit--Wigner resonance factor of the mediator $\phi$. Here $\Gamma_\phi$ is the total mediator decay width, summed over the open weak components and charge-conjugate channels. For $m_L\simeq 0$, Eq.~\eqref{eq:sigma-ann} becomes
\begin{equation}
\sigma v_{\rm lab}^{(1)}
=
\frac{y_\chi^2\lambda_L^2}{32\pi}
\frac{(s-m_\Psi^2)^2}{(s-2m_\chi^2)\left[(s-m_\phi^2)^2+m_\phi^2\Gamma_\phi^2\right]} .
\end{equation}
Using
\begin{equation}
s \simeq 4m_\chi^2 + m_\chi^2 v^2 ,
\qquad
r \equiv \frac{m_\Psi}{m_\chi} ,
\qquad
m_\phi^2 = \frac{4m_\chi^2}{1+\delta} ,
\end{equation}
for the same single channel one finds
\begin{equation}
\sigma v \simeq \sigma_0 \left( 1 + C v^2 + {\cal O}(v^4) \right) ,
\end{equation}
with
\begin{equation}
\sigma_0 =
\frac{y^2\lambda_L^2}{64\pi m_\chi^2}
\frac{(4-r^2)^2}
{\left(\frac{4\delta}{1+\delta}\right)^2+\bar\Gamma_\phi^2} ,
\qquad
\bar\Gamma_\phi \equiv \frac{m_\phi\Gamma_\phi}{m_\chi^2} = \frac{4}{1+\delta}\frac{\Gamma_\phi}{m_\phi} ,
\label{eq:sigma0}
\end{equation}
and
\begin{equation}
C  =
\frac{r^2}{2(4-r^2)}
-\frac{2B}{B^2+\bar\Gamma_\phi^2}  ,
\qquad
B \equiv \frac{4\delta}{1+\delta} .
\label{eq:Ccoeff}
\end{equation}
For most points with $C<0$, the lower dark matter temperature after kinetic decoupling enhances annihilation, giving $\Omega_\chi^{\rm cBE} h^2/\Omega_\chi^{\rm nBE}h^2<1$. Points with $C>0$ generally show the opposite trend. The magnitude of $C$ alone does not determine the relic correction. When $x_{\rm kd}$ approaches $x_{\rm cd}$, little time remains for residual annihilation, so the cBE correction is small. The resonant region also departs from a simple correlation with $|C|$. For narrow resonances such as BP1, preferential annihilation of low velocity dark matter after kinetic decoupling temporarily raises $T_\chi$. This heating caused by annihilation weakens the initial enhancement of the low velocity rate and can limit the final relic correction even when $|C|$ is large. Broader resonances behave differently. Since $C$ captures only the local dependence of $\sigma v$ as $v\to0$, a larger mediator width can smooth the Breit--Wigner variation across the thermal velocity range, making the thermally averaged rate less sensitive to $T_\chi$ than $|C|$ alone suggests. The points in Fig.~4 therefore reflect the combined effects of $C$, residual annihilation after kinetic decoupling, and the thermal evolution of dark matter.

\begin{figure}[htbp]
	\begin{center}
		 
		\includegraphics[width=8cm,height=6cm]{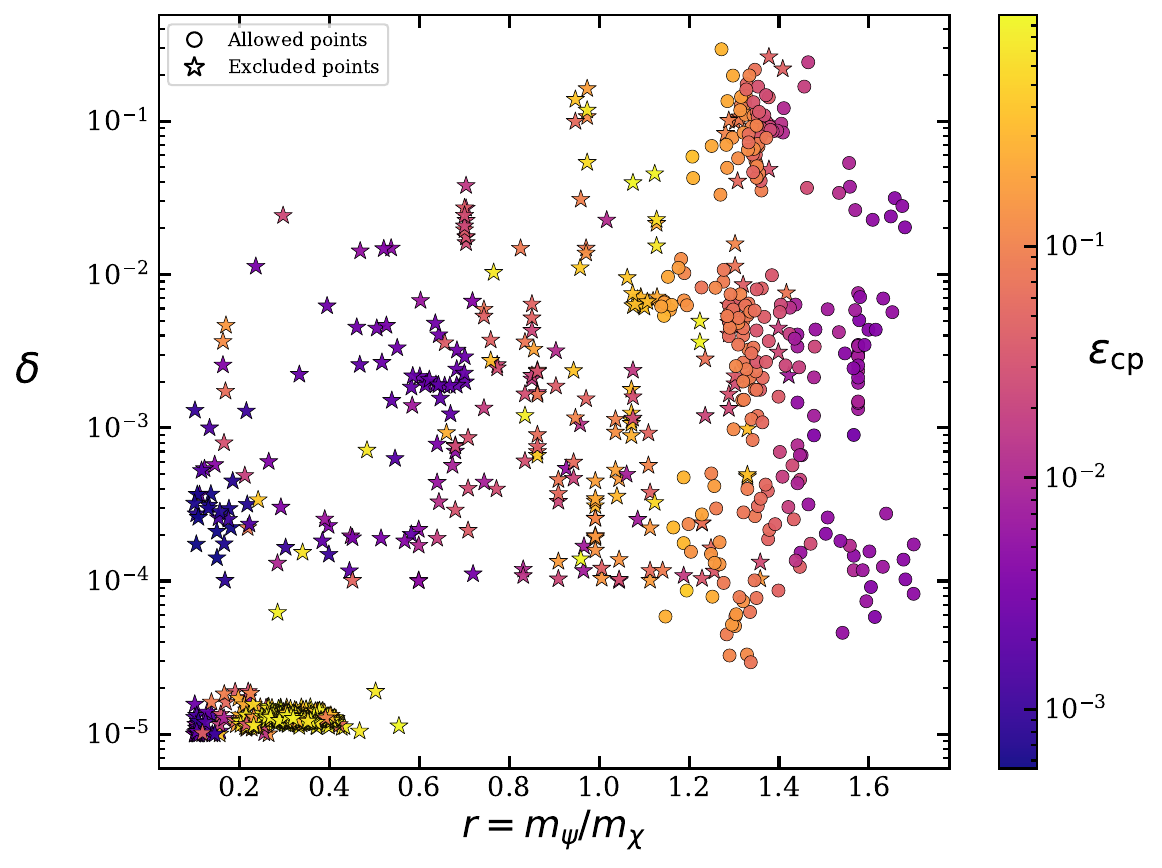}
        \includegraphics[width=8cm,height=6cm]{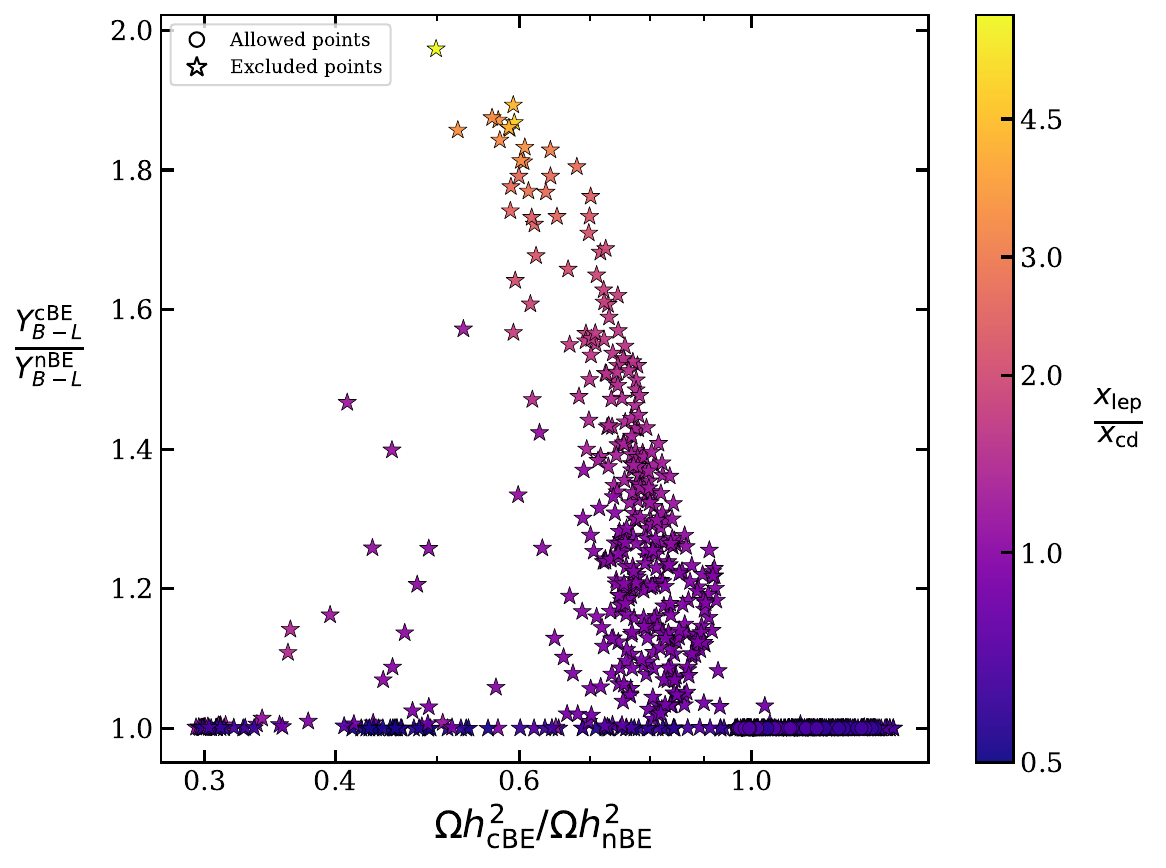}
	\end{center}
\caption{Left panel: scan points in the $r$--$\delta$ plane, colored by $\epsilon_{\rm CP}$. Right panel: asymmetry ratio $Y_{B-L}^{\rm cBE}/Y_{B-L}^{\rm nBE}$ versus ratio of relic densities, $\Omega_\chi^{\rm cBE}h^2/\Omega_\chi^{\rm nBE}h^2$, with the color scale indicating $x_{\rm lep}/x_{\rm cd}$. The asymmetry ratio is evaluated at the final evolution time.
}
	\vspace{-0.5cm}
	\label{fig:epsilon_ratio} 
\end{figure}

To further illustrate the leptogenesis behavior, the left panel of Fig.~\ref{fig:epsilon_ratio} shows that the required CP asymmetry decreases toward larger $r$ and smaller positive $\delta$. Increasing $r$ suppresses washout processes involving the heavier $\Psi$ state, allowing a larger fraction of the generated asymmetry to survive and reducing the value of $\epsilon_{\rm CP}$ needed to reproduce the observed baryon abundance. Moving closer to resonance has a similar effect through the annihilation sector. The enhanced annihilation rate allows the relic density to be obtained with smaller couplings, which further weakens washout and lowers the required $\epsilon_{\rm CP}$. The surviving region is therefore concentrated mainly at $r\gtrsim1$, where the heavier $\Psi$ state both favors early kinetic decoupling and suppresses the washout of the generated lepton asymmetry.

The right panel of Fig.~\ref{fig:epsilon_ratio} compares the asymmetry ratio with the ratio of relic densities. Most viable points remain tightly clustered around $Y_{B-L}^{\mathrm{cBE}}/Y_{B-L}^{\mathrm{nBE}}=1$, even when the relic abundance receives a sizable kinetic correction. For most of these points, $x_{\mathrm{lep}}/x_{\mathrm{cd}}<1$, so the lepton asymmetry is already close to its final value before or around chemical decoupling. However, the largest kinetic corrections generally develop after most of the asymmetry has been generated, leaving the final asymmetry nearly unchanged. When $x_{\mathrm{lep}}/x_{\mathrm{cd}}>1$, leptogenesis continues through chemical decoupling and the source becomes sensitive to both the temperature dependent annihilation rate and the evolving dark matter abundance $Y_\chi$. The resulting feedback can enhance the accumulated $B-L$ asymmetry and produce $Y_{B-L}^{\mathrm{cBE}}/Y_{B-L}^{\mathrm{nBE}}>1$. The largest deviations are found mainly among the excluded points, where the ratio reaches about $1.9734$. Within the viable parameter space, however, $x_{\mathrm{lep}}<x_{\mathrm{cd}}$ is typical and the final asymmetry remains essentially unchanged.

\section{Indirect Detection}
\label{sec:indirect}

\begin{figure}[htbp]
	\begin{center}
		 
		\includegraphics[width=7.5cm,height=7cm]{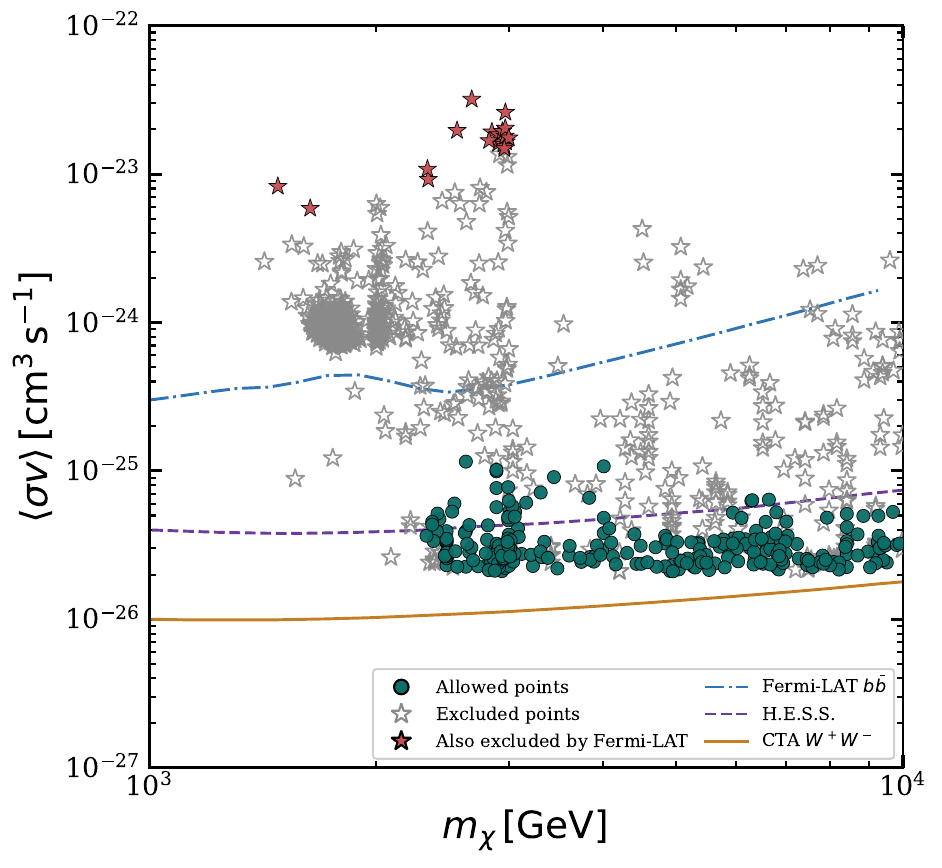}
	\end{center}
	\caption{
Effective present-day annihilation rate, \(\langle\sigma v\rangle_{\rm eff}\), as a function of \(m_\chi\). The Fermi-LAT \(b\bar b\), H.E.S.S. \(W^+W^-\), and projected CTA \(W^+W^-\) limits are shown for comparison. Grey open stars denote excluded points, while brick-red filled stars denote points also excluded by the Fermi-LAT likelihood.
}
	\vspace{-0.5cm}
	\label{fig:ID}
\end{figure}

In the interaction set considered here, the pseudoscalar mediators have no tree level coupling to quarks. Dark matter scattering off nuclei is therefore loop suppressed in the interactions considered here, and we focus on indirect detection.
The differential flux from a given target is conventionally written as~\cite{Bergstrom:1997fj,Bertone:2004pz,Cirelli:2010xx}  
\begin{equation}
\frac{{\rm d}\Phi}{{\rm d}E}
  = \frac{1}{4\pi}\,
    \frac{\langle\sigma v\rangle}{4m_{\chi}^{2}}\,
    \frac{{\rm d}N}{{\rm d}E}\,
    \int_{\Delta\Omega}{\rm d}\Omega
    \int \rho_{\rm DM}^{2}(s)\,{\rm d}s,
\label{eq:indirect_flux}
\end{equation}
Here \(\rho_{\rm DM}\) denotes the total dark matter density, while \(\langle\sigma v\rangle\) is the velocity-averaged annihilation rate. Figure~\ref{fig:ID} uses \(\langle\sigma v\rangle_{\rm eff}=\langle\sigma v\rangle/2\). The spectrum \(dN/dE\) denotes the photon yield per annihilation, and the line-of-sight integral defines the astrophysical \(J\) factor.

In the present framework, observable final states originate from the channels
$\chi\chi\to L\bar\Psi$. The heavy lepton $\Psi$ subsequently decays through
$\Psi \to Wn_R$, $Zn_R$, and $hn_R$, depending on the charged or neutral component.
The primary lepton and the subsequent decays of $W$, $Z$, and $h$ produce photons relevant for searches by Fermi-LAT, H.E.S.S., and CTA~\cite{McDaniel:2023bju,HESS:2022ygk,CTA:2020qlo}.
Because the stable light singlet $n_R$ carries away part of the annihilation energy,
the visible signal is described by the cascade spectrum. For each scan point, we use the photon spectrum generated with
\textsc{Pythia}~\cite{Bierlich:2022pythia83} together with the effective annihilation rate in the
Fermi-LAT likelihood.

Using the public energy binned likelihood for dwarf spheroidal galaxies, including the Pythia cascade spectrum and the $J$-factor uncertainty for each target, we find that the allowed parameter region remains consistent with the Fermi-LAT constraints at 95\% confidence. The published Fermi-LAT $b\bar b$ limit serves as a reference for continuum gamma-ray sensitivity. In the TeV range, the H.E.S.S. and CTA $W^+W^-$ curves provide current and projected sensitivity benchmarks for the cascade $W/Z/h+n_R$ final states considered here.

In Fig.~\ref{fig:ID}, we show the effective present-day annihilation rate, $\langle\sigma v\rangle_{\rm eff}$, at points satisfying the relic density requirement. Away from resonance, the surviving points cluster around the canonical thermal scale, $\langle\sigma v\rangle_{\rm eff}\sim 2\times10^{-26}\,\mathrm{cm}^3\,\mathrm{s}^{-1}$. Near resonance, the lower dark matter velocity at the present epoch brings annihilation closer to the Breit--Wigner pole and enhances the rate, with some excluded points reaching $3.1914\times10^{-23}\,\mathrm{cm}^3\,\mathrm{s}^{-1}$.

\section{Conclusion}
\label{sec:conclusion}

We studied a minimal $Z_4$-symmetric framework in which a singlet Dirac fermion $\chi$ constitutes dark matter, whose annihilations generate a lepton asymmetry through WIMP leptogenesis. The dark sector contains a vectorlike lepton doublet $\Psi$ and two pseudoscalar mediators $\phi_1$ and $\phi_2$. Dark matter annihilation proceeds through the off-shell $s$-channel mediator, while kinetic equilibrium is maintained predominantly by inelastic conversion $\chi L\leftrightarrow\bar\chi\Psi$ in the parameter region considered. Near resonance, the suppressed thermal abundance of the heavier mediator favors inelastic conversion over elastic scattering as the dominant source of momentum exchange. The finite mass gap suppresses inelastic conversion as the Universe cools, leading naturally to early kinetic decoupling. We therefore treat the conversion with the full inelastic collision operator and define the kinetic decoupling point through the total momentum relaxation rate, $\gamma_{\rm kin}=\gamma_{\rm conv}+\gamma_{\chi\phi}$. The resulting coupled Boltzmann system tracks the dark matter abundance, its kinetic temperature, and the lepton asymmetry, from which the visible sector $B-L$ asymmetry is reconstructed.

Our benchmark analysis shows that the impact of kinetic decoupling is governed by the relative ordering of $x_{\rm kd}$, $x_{\rm cd}$, and $x_{\rm lep}$, together with the velocity dependence of the annihilation rate. Resonance effects provide an additional source of early decoupling through the smaller couplings required by the relic density, while the resulting relic correction depends on how strongly the annihilation rate responds to the departure of $T_\chi$ from $T$. The effect on leptogenesis is much weaker in the viable region. For most accepted points, $x_{\rm lep}<x_{\rm cd}$, so the asymmetry is already close to its final value before the strongest kinetic correction develops. When $x_{\rm lep}>x_{\rm cd}$, the source remains sensitive to both the modified annihilation rate and the evolving dark matter abundance, allowing kinetic decoupling to feed back into $Y_{B-L}$. Such behavior is visible among the excluded points, where the late-time $Y_{B-L}^{\rm cBE}/Y_{B-L}^{\rm nBE}$ ratio can reach about 1.9734. The surviving parameter space is also consistent with the current Fermi-LAT dwarf spheroidal constraints.

The scan confirms that kinetic decoupling generally precedes chemical decoupling in the viable region and can modify the relic abundance in either direction, depending on the velocity dependence of the annihilation rate. The main ordering patterns are summarized in Table II. A visible relic correction requires an appreciable departure from kinetic equilibrium while annihilation is still active. The condition $x_{\rm kd}\lesssim x_{\rm cd}$ is a useful indicator but is not necessary near a strongly velocity dependent resonance. When $x_{\rm kd}\gg x_{\rm cd}$, both corrections generally become negligible. However, the ordering of these timescales alone does not determine the magnitude or direction of the relic abundance correction. The coefficient $C$ characterizes the local low-velocity behavior of $\sigma v\simeq\sigma_0(1+C v^2)$. For $T_\chi<T$, a negative $C$ generally enhances annihilation and reduces the relic abundance, while a positive $C$ tends to have the opposite effect. The final relic abundance correction also depends on residual annihilation after kinetic decoupling, annihilation-induced heating, and the mediator width. These effects together determine the final relic abundance, which cannot be inferred from $C$ alone.
\begin{table}[H]
\centering
\begin{tabular}{c c c}
\hline\hline
Ordering pattern & relic density shift & lepton asymmetry shift \\
\hline
$x_{\rm kd} \lesssim x_{\rm cd}, \quad x_{\rm cd}<x_{\rm lep}$
 & visible & visible  \\
$x_{\rm kd} \lesssim x_{\rm cd}, \quad x_{\rm lep}<x_{\rm cd}$ & visible & negligible \\
$x_{\rm kd} \gg x_{\rm cd}$ & negligible & negligible \\
\hline\hline
\end{tabular}
\caption{Qualitative summary of the expected size of cBE--nBE differences according to the relative ordering of $x_{\mathrm{kd}}$, $x_{\mathrm{cd}}$, and $x_{\mathrm{lep}}$.}
\label{tab:ordering_summary}
\end{table}

\appendix
\section{Spectator Relations and Thermal Averages}
\label{app:spectator}
For a relativistic lepton with a Fermi--Dirac distribution and $|\mu_L|/T\ll1$, the susceptibility of one weak component is $T^2/6$~\cite{Nardi:2006fx}. With the photon yield defined in the main text,
\begin{equation}
Y_{\Delta L_a}=\frac{T^2\mu_L}{6s},\qquad
Y_\gamma=\frac{2\zeta(3)T^3}{\pi^2s},\qquad
\frac{\mu_L}{T}=\frac{12\zeta(3)}{\pi^2}\frac{Y_{\Delta L_a}}{Y_\gamma}.
\label{eq:fd_lepton_normalization}
\end{equation}
For $\Delta\mu=\mu_L-\mu_\Psi=\xi\mu_L$ and $Y_{\Delta L}=2Y_{\Delta L_a}=-\eta Y_{B-L}^{\rm SM}$, this implies
\begin{equation}
\frac{\Delta\mu}{T}
=\frac{12\zeta(3)}{\pi^2}\xi\frac{Y_{\Delta L_a}}{Y_\gamma}
=\frac{6\zeta(3)}{\pi^2}\xi\frac{Y_{\Delta L}}{Y_\gamma}
=-\frac{6\zeta(3)}{\pi^2}\xi\eta\frac{Y_{B-L}^{\rm SM}}{Y_\gamma}.
\label{eq:fd_backreaction}
\end{equation}
Using Eq.~\eqref{eq:fd_backreaction}, the linearized forward-minus-conjugate reaction flux is proportional to $2\Delta\mu/T$, which gives
\begin{equation}
c_{\rm FD}=\frac{24\zeta(3)}{\pi^2},\qquad
\left.\frac{dY_{\Delta L_a}}{dx}\right|_{\rm damping}
=-\frac{c_{\rm FD}\eta\xi}{x\widetilde H}\,
\mathcal W_{\Delta L_a}Y_{\Delta L_a}.
\label{eq:fd_damping}
\end{equation}

The two spectator limits in Eq.~\eqref{eq:visible_bl} follow from the chemical-potential network of one active lepton flavor and two unsourced flavors, with all SM Yukawa interactions and electroweak sphalerons in equilibrium~\cite{Harvey:1990qw,Nardi:2006fx}. Set $q=\mu_q$, $h=\mu_H$, $\ell=\mu_{L_1}$, $m=\mu_{L_2}=\mu_{L_3}$, $p=\mu_\Psi$, and $\nu=\mu_{n_R}$, where $H$ has hypercharge $+1/2$. Define

\begin{equation}
\chi_0=\frac{T^2}{6},\qquad
\kappa_\Psi(T)=\frac{1}{\chi_0}\frac{g_\Psi}{\pi^2T}
\int_0^\infty dk\,k^2 f_\Psi^{\rm eq}(1-f_\Psi^{\rm eq}),
\qquad g_\Psi=4.
\label{eq:psi_susceptibility}
\end{equation}

Here $\kappa_\Psi$ is the susceptibility of the full Dirac doublet, with $g_\Psi=4$ particle degrees of freedom and $f_\Psi^{\rm eq}=[e^{\sqrt{k^2+m_\Psi^2}/T}+1]^{-1}$. The sector charge densities are $A_\Psi=\kappa_\Psi\chi_0p$ and $A_n=\chi_0\nu$ in thermal equilibrium, while $Y_{\Delta L}=2\chi_0\ell/s$. The relativistic and nonrelativistic limits are $\kappa_\Psi\to4$ and $\kappa_\Psi\to0$, respectively.

Yukawa equilibrium gives $\mu_u=q+h$, $\mu_d=q-h$, and $\mu_{e_i}=\mu_{L_i}-h$. Sphaleron equilibrium, hypercharge neutrality, and the unsourced charges $B/3-L_2=B/3-L_3=0$ then give

\begin{align}
9q+\ell+2m&=0,\qquad
6q-2(\ell+2m)+14h-\frac{\kappa_\Psi}{2}p=0,\nonumber\\
4q-3m+h&=0,\qquad
B-L_{\rm SM}=\chi_0(12q-3\ell-6m+3h).
\label{eq:spectator_constraints}
\end{align}

Here $B=12\chi_0q$ and $L_i$ are charge densities, with $L_i$ including the left-handed doublet and right-handed charged lepton. In the post-decoupling limit, the transfer equilibrium condition is removed and prompt sequestration requires $|p/\ell|\ll1$. Solving Eq.~\eqref{eq:spectator_constraints} with $\xi=1-p/\ell$ and $Y_{\Delta L}=-\eta Y_{B-L}^{\rm SM}$ gives
\begin{equation}
\xi_{\rm post}\simeq1,\qquad
\eta_{\rm post}=\frac{442}{711},\qquad
c_{B,\rm post}=\frac{28}{79}.
\end{equation}
This limit requires the relaxation of the $\Psi$ asymmetry to be faster than its production, with $x_{n,\rm dec}<x_{\rm w}$ providing a timing diagnostic.

While transfer remains in chemical equilibrium, the vertex $\lambda_n\overline{\Psi_L}\widetilde H n_R$ imposes $\nu=p+h$, since $\mu_{\widetilde H}=-h$. The conserved active charge $B/3-L_1-A_\Psi-A_n=0$ supplies the remaining constraint. Eliminating $m$ and $\nu$ gives

\begin{align}
35q+3\ell+2h&=0,\qquad
24q+14h-\frac{\kappa_\Psi}{2}p=0,\nonumber\\
4q-3\ell-(1+\kappa_\Psi)p&=0,\qquad
D\equiv221+223\kappa_\Psi,\nonumber\\
\frac{p}{\ell}&=-\frac{747}{D},\nonumber\\
B-L_{\rm SM}&=-\frac{9(316+329\kappa_\Psi)}{4D}\chi_0\ell.
\label{eq:spectator_eq_solution}
\end{align}

The definitions then give
\begin{equation}
\begin{gathered}
\xi_{\rm eq}=\frac{968+223\kappa_\Psi}{D},\qquad
\eta_{\rm eq}=\frac{8D}{9(316+329\kappa_\Psi)},\\
c_{B,\rm eq}=\frac{8(14+13\kappa_\Psi)}{316+329\kappa_\Psi}.
\end{gathered}
\end{equation}
The equilibrium limit applies for $x_{n,\rm dec}>x_{\rm lep}$ when transfer and the bath distributions maintain chemical equilibrium. The final sphaleron conversion uses the field content at $T_\ast$.

Since $Y_{\Delta L_a}=-(\eta/2)Y_{B-L}^{\rm SM}$, differentiating the reconstruction gives
\begin{align}
\frac{dY_{\Delta L_a}}{dx}
&=-\frac{\eta}{2}\frac{dY_{B-L}^{\rm SM}}{dx}
+\frac{d\ln\eta}{dx}Y_{\Delta L_a},\nonumber\\
\frac{d\ln\eta_{\rm eq}}{dx}
&=\left[\frac{223}{221+223\kappa_\Psi}
-\frac{329}{316+329\kappa_\Psi}\right]\frac{d\kappa_\Psi}{dx}.
\end{align}
The last term of Eq.~\eqref{eq:BE_asymmetry} vanishes for constant post-decoupling $\eta$, and for $x_{\rm w}<x_{n,\rm dec}<x_{\rm lep}$ the equation is integrated piecewise in the instantaneous-decoupling approximation, using equilibrium coefficients before $x_{n,\rm dec}$ and post-decoupling coefficients afterward. At the switch, $Y_{B-L}^{\rm SM}$ is continuous and $Y_{\Delta L_a}=-(\eta/2)Y_{B-L}^{\rm SM}$ is remapped according to
\begin{equation*}
Y_{\Delta L_a}(x_{n,\rm dec}^{+})
=\frac{\eta_{\rm post}}{\eta_{\rm eq}(x_{n,\rm dec})}
Y_{\Delta L_a}(x_{n,\rm dec}^{-}).
\end{equation*}

The heavy-doublet asymmetry $Y_{\Delta\Psi}=2Y_{\Delta\Psi_a}$ is reconstructed as
\begin{equation}
Y_{\Delta\Psi}
=\begin{cases}
0, & {\rm post},\\
\displaystyle-\frac{747\kappa_\Psi}{D}Y_{\Delta L_a}, & {\rm eq}.
\end{cases}
\label{eq:psi_asymmetry_reconstruction}
\end{equation}
The perturbative interactions conserve fermion number, whereas SM sphalerons conserve the total combination $B-F$. The corresponding Boltzmann equations therefore satisfy
\begin{equation}
\frac{d\mathcal Q_{\rm tot}}{dx}=0,
\qquad
\mathcal Q_{\rm tot}\equiv Y_{B-L}^{\rm SM}-Y_{\Delta\Psi}-Y_{\Delta n}.
\end{equation}
For the charge symmetric initial condition $\mathcal Q_{\rm tot}=0$, this gives
\begin{equation}
Y_{\Delta n}=Y_{B-L}^{\rm SM}-Y_{\Delta\Psi}.
\label{eq:n_asymmetry_relation}
\end{equation}
Thus the asymmetry stored in the $n_R$ sector is fixed by charge conservation and need not be evolved independently.
Using the continuity equation $\dot{\rho}_n+4H\rho_n=Q_n$ and entropy conservation, $s a^3=\mathrm{const.}$, we define the comoving energy density
\begin{equation}
\mathcal E_n(x)\equiv\frac{\rho_n(x)}{s(x)^{4/3}},
\qquad
\frac{d\mathcal E_n}{dx}
=\frac{Q_n(x)}{x\,\widetilde H(x)\,s(x)^{4/3}},
\label{eq:nr-radiation-evolution}
\end{equation}
Here $Q_n$ denotes the energy-injection rate per unit physical volume into the $n_R$ sector. We sum over the open decay channels, with $X=H$ above the electroweak crossover and $X=W^\pm,Z,h$ below it. For the equilibrium-parent estimate, the energy-injection rate and the daughter energy in the parent rest frame are
\begin{align}
Q_n^{\rm eq}(T)
&=\sum_{i,X}
n_i^{\rm eq}(T)\,\Gamma_{i\to Xn_R}\,E_{iX}^{*},
\\
E_{iX}^{*}
&=\frac{M_i^2-m_X^2}{2M_i}.
\label{eq:nr-equilibrium-injection}
\end{align}
Charge-conjugate decays are included. With $g_{n_R}=2$ denoting the total number of degrees of freedom of $n_R$ and $\bar n_R$, the late comoving energy and its ratio to the thermal component at decoupling are given by
\begin{align}
\mathcal E_n^{\rm late}
&=\int_{x_{n,\rm dec}}^{x_{\max}}
\frac{Q_n^{\rm eq}(x)}{x\widetilde H(x)s^{4/3}(x)}\,dx,
\\
\mathcal E_n^{\rm th}(x_{n,\rm dec})
&=\frac{\rho_n^{\rm th}(T_{n,\rm dec})}{s^{4/3}(x_{n,\rm dec})},
\qquad
\rho_n^{\rm th}(T_{n,\rm dec})
=\frac{7}{8}\frac{\pi^2}{30}g_{n_R}T_{n,\rm dec}^4,
\\
R_{\rm late}
&=\frac{\mathcal E_n^{\rm late}}{\mathcal E_n^{\rm th}(x_{n,\rm dec})},
\qquad
\Delta N_{\rm eff}^{\rm total}
=\left(1+R_{\rm late}\right)\Delta N_{\rm eff}^{\rm th}.
\label{eq:nr-late-injection-neff}
\end{align}

\phantomsection\label{app:thermal_avg}
The conventional thermal average is defined as
\begin{equation}
\langle\sigma v\rangle \;\equiv\;
\frac{g_\chi^{2}}{\bigl(n_\chi^{\rm eq}\bigr)^2}
\int\!\frac{d^{3}p}{(2\pi)^{3}}
\frac{d^{3}\tilde p}{(2\pi)^{3}}\,
\sigma v_{\chi\chi\to L\bar\Psi}^{(1)}\,
f_{\chi,{\rm eq}}({\bf p})\,
f_{\chi,{\rm eq}}(\tilde{\bf p}),
\label{eq:sigmav-def}
\end{equation}
where $n_\chi^{\rm eq}=g_\chi m_\chi^{2}T K_{2}(m_\chi/T)/(2\pi^{2})$. 
Carrying out the phase space integration one obtains the familiar expression~\cite{Binder:2017rgn}
\begin{equation}
\langle\sigma v\rangle
= \int_{1}^{\infty}\!d\tilde s\;
\sigma_{\chi\chi\to L\bar\Psi}^{(1)}\,v_{\rm lab}\,
\frac{2m_\chi \sqrt{\tilde s-1}\,(2\tilde s-1)\,
K_{1}\!\left(\tfrac{2\sqrt{\tilde s}m_\chi}{T}\right)}
{T\,K_{2}^{2}(m_\chi/T)} .
\label{eq:sigmav}
\end{equation}

The temperature weighted thermal average entering the Boltzmann hierarchy is~\cite{Binder:2017rgn},
\begin{equation}
\langle\sigma v\rangle_{2}
\;\equiv\;
\frac{g_\chi^{2}}{T\,\bigl(n_\chi^{\rm eq}\bigr)^2}
\int\!\frac{d^{3}p\,d^{3}\tilde p}{(2\pi)^{6}}
\frac{p^{2}}{3E}\,
\sigma v_{\chi\chi\to L\bar\Psi}^{(1)}\,
f_{\chi,{\rm eq}}({\bf p})\,
f_{\chi,{\rm eq}}(\tilde{\bf p}) ,
\label{eq:sigmav2-def}
\end{equation}
which can be expressed as
\begin{align}
\langle\sigma v\rangle_{2}
&= \int_{1}^{\infty}\! d\tilde s\;
\sigma_{\chi\chi\to L\bar\Psi}^{(1)}\,v_{\rm lab}\,
\frac{4\tilde s(2\tilde s-1)x^{3}}{3K_{2}^{2}(x)}
\int_{1}^{\infty}\! d\epsilon_{+}\,
e^{-2\sqrt{\tilde s}\,x\,\epsilon_{+}}
\bigg[\epsilon_{+}\sqrt{(\tilde s-1)(\epsilon_{+}^{2}-1)}
\nonumber\\
&\hspace{2cm}
+\;\frac{1}{2\sqrt{\tilde s}}\,
\log\!\left(\frac{\sqrt{\tilde s}\,\epsilon_{+}
-\sqrt{(\tilde s-1)(\epsilon_{+}^{2}-1)}}
{\sqrt{\tilde s}\,\epsilon_{+}
+\sqrt{(\tilde s-1)(\epsilon_{+}^{2}-1)}}\right)\bigg],
\label{eq:sigmav2}
\end{align}

\begin{acknowledgments}
We thank Cheng-Feng Cai, Zhi-Long Han, Yi-Lei Tang and Zhao-Huan Yu for discussions. This work is supported by the National Natural Science Foundation of China (NSFC) under Grant No.~12275367, the Fundamental Research Funds for the Central Universities, and the Sun Yat-Sen University Science Foundation.
\end{acknowledgments}

\bibliographystyle{apsrev4-1}
\bibliography{references}

\begin{thebibliography}{57}%
\makeatletter
\providecommand \@ifxundefined [1]{%
 \@ifx{#1\undefined}
}%
\providecommand \@ifnum [1]{%
 \ifnum #1\expandafter \@firstoftwo
 \else \expandafter \@secondoftwo
 \fi
}%
\providecommand \@ifx [1]{%
 \ifx #1\expandafter \@firstoftwo
 \else \expandafter \@secondoftwo
 \fi
}%
\providecommand \natexlab [1]{#1}%
\providecommand \enquote  [1]{``#1''}%
\providecommand \bibnamefont  [1]{#1}%
\providecommand \bibfnamefont [1]{#1}%
\providecommand \citenamefont [1]{#1}%
\providecommand \href@noop [0]{\@secondoftwo}%
\providecommand \href [0]{\begingroup \@sanitize@url \@href}%
\providecommand \@href[1]{\@@startlink{#1}\@@href}%
\providecommand \@@href[1]{\endgroup#1\@@endlink}%
\providecommand \@sanitize@url [0]{\catcode `\\12\catcode `\$12\catcode
  `\&12\catcode `\#12\catcode `\^12\catcode `\_12\catcode `\%12\relax}%
\providecommand \@@startlink[1]{}%
\providecommand \@@endlink[0]{}%
\providecommand \url  [0]{\begingroup\@sanitize@url \@url }%
\providecommand \@url [1]{\endgroup\@href {#1}{\urlprefix }}%
\providecommand \urlprefix  [0]{URL }%
\providecommand \Eprint [0]{\href }%
\providecommand \doibase [0]{http://dx.doi.org/}%
\providecommand \selectlanguage [0]{\@gobble}%
\providecommand \bibinfo  [0]{\@secondoftwo}%
\providecommand \bibfield  [0]{\@secondoftwo}%
\providecommand \translation [1]{[#1]}%
\providecommand \BibitemOpen [0]{}%
\providecommand \bibitemStop [0]{}%
\providecommand \bibitemNoStop [0]{.\EOS\space}%
\providecommand \EOS [0]{\spacefactor3000\relax}%
\providecommand \BibitemShut  [1]{\csname bibitem#1\endcsname}%
\let\auto@bib@innerbib\@empty
\bibitem [{\citenamefont {Bertone}\ \emph {et~al.}(2005)\citenamefont
  {Bertone}, \citenamefont {Hooper},\ and\ \citenamefont
  {Silk}}]{Bertone:2004pz}%
  \BibitemOpen
  \bibfield  {author} {\bibinfo {author} {\bibfnamefont {G.}~\bibnamefont
  {Bertone}}, \bibinfo {author} {\bibfnamefont {D.}~\bibnamefont {Hooper}}, \
  and\ \bibinfo {author} {\bibfnamefont {J.}~\bibnamefont {Silk}},\ }\href
  {\doibase 10.1016/j.physrep.2004.08.031} {\bibfield  {journal} {\bibinfo
  {journal} {Phys. Rept.}\ }\textbf {\bibinfo {volume} {405}},\ \bibinfo
  {pages} {279} (\bibinfo {year} {2005})},\ \Eprint
  {http://arxiv.org/abs/hep-ph/0404175} {hep-ph/0404175 [hep-ph]} \BibitemShut
  {NoStop}%
\bibitem [{\citenamefont {Aghanim}\ and\ \citenamefont {{Planck
  Collaboration}}(2020)}]{Planck:2018vyg}%
  \BibitemOpen
  \bibfield  {author} {\bibinfo {author} {\bibfnamefont {N.}~\bibnamefont
  {Aghanim}}\ and\ \bibinfo {author} {\bibnamefont {{Planck Collaboration}}},\
  }\href {\doibase 10.1051/0004-6361/201833910} {\bibfield  {journal} {\bibinfo
   {journal} {Astron. Astrophys.}\ }\textbf {\bibinfo {volume} {641}},\
  \bibinfo {pages} {A6} (\bibinfo {year} {2020})},\ \bibinfo {note} {[erratum:
  Astron. Astrophys. \textbf{652} (2021), C4]},\ \Eprint
  {http://arxiv.org/abs/1807.06209} {1807.06209 [astro-ph.CO]} \BibitemShut
  {NoStop}%
\bibitem [{\citenamefont {Fields}\ \emph {et~al.}(2020)\citenamefont {Fields},
  \citenamefont {Olive}, \citenamefont {Yeh},\ and\ \citenamefont
  {Young}}]{Fields:2019pfx}%
  \BibitemOpen
  \bibfield  {author} {\bibinfo {author} {\bibfnamefont {B.~D.}\ \bibnamefont
  {Fields}}, \bibinfo {author} {\bibfnamefont {K.~A.}\ \bibnamefont {Olive}},
  \bibinfo {author} {\bibfnamefont {T.~H.}\ \bibnamefont {Yeh}}, \ and\
  \bibinfo {author} {\bibfnamefont {C.}~\bibnamefont {Young}},\ }\href
  {\doibase 10.1088/1475-7516/2020/03/010} {\bibfield  {journal} {\bibinfo
  {journal} {JCAP}\ }\textbf {\bibinfo {volume} {03}},\ \bibinfo {pages} {010}
  (\bibinfo {year} {2020})},\ \bibinfo {note} {[erratum: JCAP \textbf{11}
  (2020), E02]},\ \Eprint {http://arxiv.org/abs/1912.01132} {1912.01132
  [astro-ph.CO]} \BibitemShut {NoStop}%
\bibitem [{\citenamefont {D'Onofrio}\ \emph {et~al.}(2014)\citenamefont
  {D'Onofrio}, \citenamefont {Rummukainen},\ and\ \citenamefont
  {Tranberg}}]{DOnofrio:2014rug}%
  \BibitemOpen
  \bibfield  {author} {\bibinfo {author} {\bibfnamefont {M.}~\bibnamefont
  {D'Onofrio}}, \bibinfo {author} {\bibfnamefont {K.}~\bibnamefont
  {Rummukainen}}, \ and\ \bibinfo {author} {\bibfnamefont {A.}~\bibnamefont
  {Tranberg}},\ }\href {\doibase 10.1103/PhysRevLett.113.141602} {\bibfield
  {journal} {\bibinfo  {journal} {Phys. Rev. Lett.}\ }\textbf {\bibinfo
  {volume} {113}},\ \bibinfo {pages} {141602} (\bibinfo {year} {2014})},\
  \Eprint {http://arxiv.org/abs/1404.3565} {1404.3565 [hep-ph]} \BibitemShut
  {NoStop}%
\bibitem [{\citenamefont {Fukugita}\ and\ \citenamefont
  {Yanagida}(1986)}]{Fukugita:1986hr}%
  \BibitemOpen
  \bibfield  {author} {\bibinfo {author} {\bibfnamefont {M.}~\bibnamefont
  {Fukugita}}\ and\ \bibinfo {author} {\bibfnamefont {T.}~\bibnamefont
  {Yanagida}},\ }\href {\doibase 10.1016/0370-2693(86)91126-3} {\bibfield
  {journal} {\bibinfo  {journal} {Phys. Lett. B}\ }\textbf {\bibinfo {volume}
  {174}},\ \bibinfo {pages} {45} (\bibinfo {year} {1986})}\BibitemShut
  {NoStop}%
\bibitem [{\citenamefont {Buchmuller}\ \emph {et~al.}(2005)\citenamefont
  {Buchmuller}, \citenamefont {Peccei},\ and\ \citenamefont
  {Yanagida}}]{Buchmuller:2005eh}%
  \BibitemOpen
  \bibfield  {author} {\bibinfo {author} {\bibfnamefont {W.}~\bibnamefont
  {Buchmuller}}, \bibinfo {author} {\bibfnamefont {R.~D.}\ \bibnamefont
  {Peccei}}, \ and\ \bibinfo {author} {\bibfnamefont {T.}~\bibnamefont
  {Yanagida}},\ }\href {\doibase 10.1146/annurev.nucl.55.090704.151558}
  {\bibfield  {journal} {\bibinfo  {journal} {Ann. Rev. Nucl. Part. Sci.}\
  }\textbf {\bibinfo {volume} {55}},\ \bibinfo {pages} {311} (\bibinfo {year}
  {2005})},\ \Eprint {http://arxiv.org/abs/hep-ph/0502169} {hep-ph/0502169
  [hep-ph]} \BibitemShut {NoStop}%
\bibitem [{\citenamefont {Manton}(1983)}]{Manton:1983nd}%
  \BibitemOpen
  \bibfield  {author} {\bibinfo {author} {\bibfnamefont {N.~S.}\ \bibnamefont
  {Manton}},\ }\href {\doibase 10.1103/PhysRevD.28.2019} {\bibfield  {journal}
  {\bibinfo  {journal} {Phys. Rev. D}\ }\textbf {\bibinfo {volume} {28}},\
  \bibinfo {pages} {2019} (\bibinfo {year} {1983})}\BibitemShut {NoStop}%
\bibitem [{\citenamefont {Klinkhamer}\ and\ \citenamefont
  {Manton}(1984)}]{Klinkhamer:1984di}%
  \BibitemOpen
  \bibfield  {author} {\bibinfo {author} {\bibfnamefont {F.~R.}\ \bibnamefont
  {Klinkhamer}}\ and\ \bibinfo {author} {\bibfnamefont {N.~S.}\ \bibnamefont
  {Manton}},\ }\href {\doibase 10.1103/PhysRevD.30.2212} {\bibfield  {journal}
  {\bibinfo  {journal} {Phys. Rev. D}\ }\textbf {\bibinfo {volume} {30}},\
  \bibinfo {pages} {2212} (\bibinfo {year} {1984})}\BibitemShut {NoStop}%
\bibitem [{\citenamefont {Khlebnikov}\ and\ \citenamefont
  {Shaposhnikov}(1988)}]{Khlebnikov:1988sr}%
  \BibitemOpen
  \bibfield  {author} {\bibinfo {author} {\bibfnamefont {S.~Y.}\ \bibnamefont
  {Khlebnikov}}\ and\ \bibinfo {author} {\bibfnamefont {M.~E.}\ \bibnamefont
  {Shaposhnikov}},\ }\href {\doibase 10.1016/0550-3213(88)90133-2} {\bibfield
  {journal} {\bibinfo  {journal} {Nucl. Phys. B}\ }\textbf {\bibinfo {volume}
  {308}},\ \bibinfo {pages} {885} (\bibinfo {year} {1988})}\BibitemShut
  {NoStop}%
\bibitem [{\citenamefont {Elor}\ \emph {et~al.}(2019)\citenamefont {Elor},
  \citenamefont {Escudero},\ and\ \citenamefont {Nelson}}]{Elor:2018twp}%
  \BibitemOpen
  \bibfield  {author} {\bibinfo {author} {\bibfnamefont {G.}~\bibnamefont
  {Elor}}, \bibinfo {author} {\bibfnamefont {M.}~\bibnamefont {Escudero}}, \
  and\ \bibinfo {author} {\bibfnamefont {A.}~\bibnamefont {Nelson}},\ }\href
  {\doibase 10.1103/PhysRevD.99.035031} {\bibfield  {journal} {\bibinfo
  {journal} {Phys. Rev. D}\ }\textbf {\bibinfo {volume} {99}},\ \bibinfo
  {pages} {035031} (\bibinfo {year} {2019})},\ \Eprint
  {http://arxiv.org/abs/1810.00880} {1810.00880 [hep-ph]} \BibitemShut
  {NoStop}%
\bibitem [{\citenamefont {Shuve}\ and\ \citenamefont
  {Tucker-Smith}(2020)}]{Shuve:2020evk}%
  \BibitemOpen
  \bibfield  {author} {\bibinfo {author} {\bibfnamefont {B.}~\bibnamefont
  {Shuve}}\ and\ \bibinfo {author} {\bibfnamefont {D.}~\bibnamefont
  {Tucker-Smith}},\ }\href {\doibase 10.1103/PhysRevD.101.115023} {\bibfield
  {journal} {\bibinfo  {journal} {Phys. Rev. D}\ }\textbf {\bibinfo {volume}
  {101}},\ \bibinfo {pages} {115023} (\bibinfo {year} {2020})},\ \Eprint
  {http://arxiv.org/abs/2004.00636} {2004.00636 [hep-ph]} \BibitemShut
  {NoStop}%
\bibitem [{\citenamefont {Hall}\ \emph {et~al.}(2023)\citenamefont {Hall},
  \citenamefont {Konstandin}, \citenamefont {McGehee},\ and\ \citenamefont
  {Murayama}}]{Hall:2019rld}%
  \BibitemOpen
  \bibfield  {author} {\bibinfo {author} {\bibfnamefont {E.}~\bibnamefont
  {Hall}}, \bibinfo {author} {\bibfnamefont {T.}~\bibnamefont {Konstandin}},
  \bibinfo {author} {\bibfnamefont {R.}~\bibnamefont {McGehee}}, \ and\
  \bibinfo {author} {\bibfnamefont {H.}~\bibnamefont {Murayama}},\ }\href
  {\doibase 10.1103/PhysRevD.107.055011} {\bibfield  {journal} {\bibinfo
  {journal} {Phys. Rev. D}\ }\textbf {\bibinfo {volume} {107}},\ \bibinfo
  {pages} {055011} (\bibinfo {year} {2023})},\ \Eprint
  {http://arxiv.org/abs/1911.12342} {1911.12342 [hep-ph]} \BibitemShut
  {NoStop}%
\bibitem [{\citenamefont {Carr}\ \emph {et~al.}(2021)\citenamefont {Carr},
  \citenamefont {Clesse},\ and\ \citenamefont
  {Garc{\'\i}a-Bellido}}]{Carr:2019hud}%
  \BibitemOpen
  \bibfield  {author} {\bibinfo {author} {\bibfnamefont {B.}~\bibnamefont
  {Carr}}, \bibinfo {author} {\bibfnamefont {S.}~\bibnamefont {Clesse}}, \ and\
  \bibinfo {author} {\bibfnamefont {J.}~\bibnamefont {Garc{\'\i}a-Bellido}},\
  }\href {\doibase 10.1093/mnras/staa3726} {\bibfield  {journal} {\bibinfo
  {journal} {Mon. Not. Roy. Astron. Soc.}\ }\textbf {\bibinfo {volume} {501}},\
  \bibinfo {pages} {1426} (\bibinfo {year} {2021})},\ \Eprint
  {http://arxiv.org/abs/1904.02129} {1904.02129 [astro-ph.CO]} \BibitemShut
  {NoStop}%
\bibitem [{\citenamefont {Ma}(2006)}]{Ma:2006fn}%
  \BibitemOpen
  \bibfield  {author} {\bibinfo {author} {\bibfnamefont {E.}~\bibnamefont
  {Ma}},\ }\href {\doibase 10.1142/S0217732306021141} {\bibfield  {journal}
  {\bibinfo  {journal} {Mod. Phys. Lett. A}\ }\textbf {\bibinfo {volume}
  {21}},\ \bibinfo {pages} {1777} (\bibinfo {year} {2006})},\ \Eprint
  {http://arxiv.org/abs/hep-ph/0605180} {hep-ph/0605180 [hep-ph]} \BibitemShut
  {NoStop}%
\bibitem [{\citenamefont {Falkowski}\ \emph {et~al.}(2019)\citenamefont
  {Falkowski}, \citenamefont {Kuflik}, \citenamefont {Levi},\ and\
  \citenamefont {Volansky}}]{Falkowski:2017uya}%
  \BibitemOpen
  \bibfield  {author} {\bibinfo {author} {\bibfnamefont {A.}~\bibnamefont
  {Falkowski}}, \bibinfo {author} {\bibfnamefont {E.}~\bibnamefont {Kuflik}},
  \bibinfo {author} {\bibfnamefont {N.}~\bibnamefont {Levi}}, \ and\ \bibinfo
  {author} {\bibfnamefont {T.}~\bibnamefont {Volansky}},\ }\href {\doibase
  10.1103/PhysRevD.99.015022} {\bibfield  {journal} {\bibinfo  {journal} {Phys.
  Rev. D}\ }\textbf {\bibinfo {volume} {99}},\ \bibinfo {pages} {015022}
  (\bibinfo {year} {2019})},\ \Eprint {http://arxiv.org/abs/1712.07652}
  {1712.07652 [hep-ph]} \BibitemShut {NoStop}%
\bibitem [{\citenamefont {Liu}\ \emph {et~al.}(2020)\citenamefont {Liu},
  \citenamefont {Han}, \citenamefont {Jin},\ and\ \citenamefont
  {Yang}}]{Liu:2020mxj}%
  \BibitemOpen
  \bibfield  {author} {\bibinfo {author} {\bibfnamefont {A.}~\bibnamefont
  {Liu}}, \bibinfo {author} {\bibfnamefont {Z.~L.}\ \bibnamefont {Han}},
  \bibinfo {author} {\bibfnamefont {Y.}~\bibnamefont {Jin}}, \ and\ \bibinfo
  {author} {\bibfnamefont {F.~X.}\ \bibnamefont {Yang}},\ }\href {\doibase
  10.1103/PhysRevD.101.095005} {\bibfield  {journal} {\bibinfo  {journal}
  {Phys. Rev. D}\ }\textbf {\bibinfo {volume} {101}},\ \bibinfo {pages}
  {095005} (\bibinfo {year} {2020})},\ \Eprint
  {http://arxiv.org/abs/2001.04085} {2001.04085 [hep-ph]} \BibitemShut
  {NoStop}%
\bibitem [{\citenamefont {Petraki}\ and\ \citenamefont
  {Volkas}(2013)}]{Petraki:2013wwa}%
  \BibitemOpen
  \bibfield  {author} {\bibinfo {author} {\bibfnamefont {K.}~\bibnamefont
  {Petraki}}\ and\ \bibinfo {author} {\bibfnamefont {R.~R.}\ \bibnamefont
  {Volkas}},\ }\href {\doibase 10.1142/S0217751X13300287} {\bibfield  {journal}
  {\bibinfo  {journal} {Int. J. Mod. Phys. A}\ }\textbf {\bibinfo {volume}
  {28}},\ \bibinfo {pages} {1330028} (\bibinfo {year} {2013})},\ \Eprint
  {http://arxiv.org/abs/1305.4939} {1305.4939 [hep-ph]} \BibitemShut {NoStop}%
\bibitem [{\citenamefont {Zurek}(2014)}]{Zurek:2013wia}%
  \BibitemOpen
  \bibfield  {author} {\bibinfo {author} {\bibfnamefont {K.~M.}\ \bibnamefont
  {Zurek}},\ }\href {\doibase 10.1016/j.physrep.2013.12.001} {\bibfield
  {journal} {\bibinfo  {journal} {Phys. Rept.}\ }\textbf {\bibinfo {volume}
  {537}},\ \bibinfo {pages} {91} (\bibinfo {year} {2014})},\ \Eprint
  {http://arxiv.org/abs/1308.0338} {1308.0338 [hep-ph]} \BibitemShut {NoStop}%
\bibitem [{\citenamefont {Kaplan}\ \emph {et~al.}(2009)\citenamefont {Kaplan},
  \citenamefont {Luty},\ and\ \citenamefont {Zurek}}]{Kaplan:2009ag}%
  \BibitemOpen
  \bibfield  {author} {\bibinfo {author} {\bibfnamefont {D.~E.}\ \bibnamefont
  {Kaplan}}, \bibinfo {author} {\bibfnamefont {M.~A.}\ \bibnamefont {Luty}}, \
  and\ \bibinfo {author} {\bibfnamefont {K.~M.}\ \bibnamefont {Zurek}},\ }\href
  {\doibase 10.1103/PhysRevD.79.115016} {\bibfield  {journal} {\bibinfo
  {journal} {Phys. Rev. D}\ }\textbf {\bibinfo {volume} {79}},\ \bibinfo
  {pages} {115016} (\bibinfo {year} {2009})},\ \Eprint
  {http://arxiv.org/abs/0901.4117} {0901.4117 [hep-ph]} \BibitemShut {NoStop}%
\bibitem [{\citenamefont {Cui}\ \emph {et~al.}(2012)\citenamefont {Cui},
  \citenamefont {Randall},\ and\ \citenamefont {Shuve}}]{Cui:2011ab}%
  \BibitemOpen
  \bibfield  {author} {\bibinfo {author} {\bibfnamefont {Y.}~\bibnamefont
  {Cui}}, \bibinfo {author} {\bibfnamefont {L.}~\bibnamefont {Randall}}, \ and\
  \bibinfo {author} {\bibfnamefont {B.}~\bibnamefont {Shuve}},\ }\href
  {\doibase 10.1007/JHEP04(2012)075} {\bibfield  {journal} {\bibinfo  {journal}
  {JHEP}\ }\textbf {\bibinfo {volume} {04}},\ \bibinfo {pages} {075} (\bibinfo
  {year} {2012})},\ \Eprint {http://arxiv.org/abs/1112.2704} {1112.2704
  [hep-ph]} \BibitemShut {NoStop}%
\bibitem [{\citenamefont {Bernal}\ \emph
  {et~al.}(2013{\natexlab{a}})\citenamefont {Bernal}, \citenamefont {Colucci},
  \citenamefont {Josse-Michaux}, \citenamefont {Racker},\ and\ \citenamefont
  {Ubaldi}}]{Bernal:2013bga}%
  \BibitemOpen
  \bibfield  {author} {\bibinfo {author} {\bibfnamefont {N.}~\bibnamefont
  {Bernal}}, \bibinfo {author} {\bibfnamefont {S.}~\bibnamefont {Colucci}},
  \bibinfo {author} {\bibfnamefont {F.~X.}\ \bibnamefont {Josse-Michaux}},
  \bibinfo {author} {\bibfnamefont {J.}~\bibnamefont {Racker}}, \ and\ \bibinfo
  {author} {\bibfnamefont {L.}~\bibnamefont {Ubaldi}},\ }\href {\doibase
  10.1088/1475-7516/2013/10/035} {\bibfield  {journal} {\bibinfo  {journal}
  {JCAP}\ }\textbf {\bibinfo {volume} {10}},\ \bibinfo {pages} {035} (\bibinfo
  {year} {2013}{\natexlab{a}})},\ \Eprint {http://arxiv.org/abs/1307.6878}
  {1307.6878 [hep-ph]} \BibitemShut {NoStop}%
\bibitem [{\citenamefont {Davidson}\ and\ \citenamefont
  {Ibarra}(2002)}]{Davidson:2002qv}%
  \BibitemOpen
  \bibfield  {author} {\bibinfo {author} {\bibfnamefont {S.}~\bibnamefont
  {Davidson}}\ and\ \bibinfo {author} {\bibfnamefont {A.}~\bibnamefont
  {Ibarra}},\ }\href {\doibase 10.1016/S0370-2693(02)01735-5} {\bibfield
  {journal} {\bibinfo  {journal} {Phys. Lett. B}\ }\textbf {\bibinfo {volume}
  {535}},\ \bibinfo {pages} {25} (\bibinfo {year} {2002})},\ \Eprint
  {http://arxiv.org/abs/hep-ph/0202239} {hep-ph/0202239 [hep-ph]} \BibitemShut
  {NoStop}%
\bibitem [{\citenamefont {Calabrese}\ \emph {et~al.}(2024)\citenamefont
  {Calabrese}, \citenamefont {Chianese}, \citenamefont {Gunn}, \citenamefont
  {Miele}, \citenamefont {Morisi},\ and\ \citenamefont
  {Saviano}}]{Calabrese:2023bxz}%
  \BibitemOpen
  \bibfield  {author} {\bibinfo {author} {\bibfnamefont {R.}~\bibnamefont
  {Calabrese}}, \bibinfo {author} {\bibfnamefont {M.}~\bibnamefont {Chianese}},
  \bibinfo {author} {\bibfnamefont {J.}~\bibnamefont {Gunn}}, \bibinfo {author}
  {\bibfnamefont {G.}~\bibnamefont {Miele}}, \bibinfo {author} {\bibfnamefont
  {S.}~\bibnamefont {Morisi}}, \ and\ \bibinfo {author} {\bibfnamefont
  {N.}~\bibnamefont {Saviano}},\ }\href {\doibase 10.1103/PhysRevD.109.103001}
  {\bibfield  {journal} {\bibinfo  {journal} {Phys. Rev. D}\ }\textbf {\bibinfo
  {volume} {109}},\ \bibinfo {pages} {103001} (\bibinfo {year} {2024})},\
  \Eprint {http://arxiv.org/abs/2311.13276} {2311.13276 [hep-ph]} \BibitemShut
  {NoStop}%
\bibitem [{\citenamefont {Heisig}(2024)}]{Heisig:2024conversion}%
  \BibitemOpen
  \bibfield  {author} {\bibinfo {author} {\bibfnamefont {J.}~\bibnamefont
  {Heisig}},\ }\href {\doibase 10.1103/PhysRevLett.133.191803} {\bibfield
  {journal} {\bibinfo  {journal} {Phys. Rev. Lett.}\ }\textbf {\bibinfo
  {volume} {133}},\ \bibinfo {pages} {191803} (\bibinfo {year} {2024})},\
  \Eprint {http://arxiv.org/abs/2404.12428} {arXiv:2404.12428 [hep-ph]}
  \BibitemShut {NoStop}%
\bibitem [{\citenamefont {Bernal}\ \emph
  {et~al.}(2013{\natexlab{b}})\citenamefont {Bernal}, \citenamefont
  {Josse-Michaux},\ and\ \citenamefont {Ubaldi}}]{Bernal:2012gv}%
  \BibitemOpen
  \bibfield  {author} {\bibinfo {author} {\bibfnamefont {N.}~\bibnamefont
  {Bernal}}, \bibinfo {author} {\bibfnamefont {F.~X.}\ \bibnamefont
  {Josse-Michaux}}, \ and\ \bibinfo {author} {\bibfnamefont {L.}~\bibnamefont
  {Ubaldi}},\ }\href {\doibase 10.1088/1475-7516/2013/01/034} {\bibfield
  {journal} {\bibinfo  {journal} {JCAP}\ }\textbf {\bibinfo {volume} {01}},\
  \bibinfo {pages} {034} (\bibinfo {year} {2013}{\natexlab{b}})},\ \Eprint
  {http://arxiv.org/abs/1210.0094} {1210.0094 [hep-ph]} \BibitemShut {NoStop}%
\bibitem [{\citenamefont {Mahanta}\ and\ \citenamefont
  {Borah}(2023)}]{Mahanta:2022WIMPy}%
  \BibitemOpen
  \bibfield  {author} {\bibinfo {author} {\bibfnamefont {D.}~\bibnamefont
  {Mahanta}}\ and\ \bibinfo {author} {\bibfnamefont {D.}~\bibnamefont
  {Borah}},\ }\href {\doibase 10.1088/1475-7516/2023/03/049} {\bibfield
  {journal} {\bibinfo  {journal} {JCAP}\ }\textbf {\bibinfo {volume} {03}},\
  \bibinfo {pages} {049} (\bibinfo {year} {2023})},\ \Eprint
  {http://arxiv.org/abs/2208.11295} {2208.11295 [hep-ph]} \BibitemShut
  {NoStop}%
\bibitem [{\citenamefont {Binder}\ \emph {et~al.}(2017)\citenamefont {Binder},
  \citenamefont {Bringmann}, \citenamefont {Gustafsson},\ and\ \citenamefont
  {Hryczuk}}]{Binder:2017rgn}%
  \BibitemOpen
  \bibfield  {author} {\bibinfo {author} {\bibfnamefont {T.}~\bibnamefont
  {Binder}}, \bibinfo {author} {\bibfnamefont {T.}~\bibnamefont {Bringmann}},
  \bibinfo {author} {\bibfnamefont {M.}~\bibnamefont {Gustafsson}}, \ and\
  \bibinfo {author} {\bibfnamefont {A.}~\bibnamefont {Hryczuk}},\ }\href
  {\doibase 10.1103/PhysRevD.96.115010} {\bibfield  {journal} {\bibinfo
  {journal} {Phys. Rev. D}\ }\textbf {\bibinfo {volume} {96}},\ \bibinfo
  {pages} {115010} (\bibinfo {year} {2017})},\ \bibinfo {note} {[erratum: Phys.
  Rev. D \textbf{101} (2020) no.9, 099901]},\ \Eprint
  {http://arxiv.org/abs/1706.07433} {1706.07433 [astro-ph.CO]} \BibitemShut
  {NoStop}%
\bibitem [{\citenamefont {Duch}\ and\ \citenamefont
  {Grzadkowski}(2017)}]{Duch:2017nbe}%
  \BibitemOpen
  \bibfield  {author} {\bibinfo {author} {\bibfnamefont {M.}~\bibnamefont
  {Duch}}\ and\ \bibinfo {author} {\bibfnamefont {B.}~\bibnamefont
  {Grzadkowski}},\ }\href {\doibase 10.1007/JHEP09(2017)159} {\bibfield
  {journal} {\bibinfo  {journal} {JHEP}\ }\textbf {\bibinfo {volume} {09}},\
  \bibinfo {pages} {159} (\bibinfo {year} {2017})},\ \Eprint
  {http://arxiv.org/abs/1705.10777} {1705.10777 [hep-ph]} \BibitemShut
  {NoStop}%
\bibitem [{\citenamefont {Abe}(2020)}]{Abe:2020obo}%
  \BibitemOpen
  \bibfield  {author} {\bibinfo {author} {\bibfnamefont {T.}~\bibnamefont
  {Abe}},\ }\href {\doibase 10.1103/PhysRevD.102.035018} {\bibfield  {journal}
  {\bibinfo  {journal} {Phys. Rev. D}\ }\textbf {\bibinfo {volume} {102}},\
  \bibinfo {pages} {035018} (\bibinfo {year} {2020})},\ \Eprint
  {http://arxiv.org/abs/2004.10041} {2004.10041 [hep-ph]} \BibitemShut
  {NoStop}%
\bibitem [{\citenamefont {Bi}\ \emph {et~al.}(2012)\citenamefont {Bi},
  \citenamefont {Yin},\ and\ \citenamefont {Yuan}}]{Bi:2011qm}%
  \BibitemOpen
  \bibfield  {author} {\bibinfo {author} {\bibfnamefont {X.~J.}\ \bibnamefont
  {Bi}}, \bibinfo {author} {\bibfnamefont {P.~F.}\ \bibnamefont {Yin}}, \ and\
  \bibinfo {author} {\bibfnamefont {Q.}~\bibnamefont {Yuan}},\ }\href {\doibase
  10.1103/PhysRevD.85.043526} {\bibfield  {journal} {\bibinfo  {journal} {Phys.
  Rev. D}\ }\textbf {\bibinfo {volume} {85}},\ \bibinfo {pages} {043526}
  (\bibinfo {year} {2012})},\ \Eprint {http://arxiv.org/abs/1106.6027}
  {1106.6027 [hep-ph]} \BibitemShut {NoStop}%
\bibitem [{\citenamefont {Binder}\ \emph {et~al.}(2021)\citenamefont {Binder},
  \citenamefont {Bringmann}, \citenamefont {Gustafsson},\ and\ \citenamefont
  {Hryczuk}}]{Binder:2021bmg}%
  \BibitemOpen
  \bibfield  {author} {\bibinfo {author} {\bibfnamefont {T.}~\bibnamefont
  {Binder}}, \bibinfo {author} {\bibfnamefont {T.}~\bibnamefont {Bringmann}},
  \bibinfo {author} {\bibfnamefont {M.}~\bibnamefont {Gustafsson}}, \ and\
  \bibinfo {author} {\bibfnamefont {A.}~\bibnamefont {Hryczuk}},\ }\href
  {\doibase 10.1140/epjc/s10052-021-09357-5} {\bibfield  {journal} {\bibinfo
  {journal} {Eur. Phys. J. C}\ }\textbf {\bibinfo {volume} {81}},\ \bibinfo
  {pages} {577} (\bibinfo {year} {2021})},\ \Eprint
  {http://arxiv.org/abs/2103.01944} {2103.01944 [hep-ph]} \BibitemShut
  {NoStop}%
\bibitem [{\citenamefont {Liu}\ \emph {et~al.}(2023)\citenamefont {Liu},
  \citenamefont {Liu},\ and\ \citenamefont {Zhu}}]{Liu:2023kat}%
  \BibitemOpen
  \bibfield  {author} {\bibinfo {author} {\bibfnamefont {Y.}~\bibnamefont
  {Liu}}, \bibinfo {author} {\bibfnamefont {X.}~\bibnamefont {Liu}}, \ and\
  \bibinfo {author} {\bibfnamefont {B.}~\bibnamefont {Zhu}},\ }\href {\doibase
  10.1103/PhysRevD.107.115009} {\bibfield  {journal} {\bibinfo  {journal}
  {Phys. Rev. D}\ }\textbf {\bibinfo {volume} {107}},\ \bibinfo {pages}
  {115009} (\bibinfo {year} {2023})},\ \Eprint
  {http://arxiv.org/abs/2301.12199} {2301.12199 [hep-ph]} \BibitemShut
  {NoStop}%
\bibitem [{\citenamefont {Aboubrahim}\ \emph {et~al.}(2023)\citenamefont
  {Aboubrahim}, \citenamefont {Klasen},\ and\ \citenamefont
  {Wiggering}}]{Aboubrahim:2023yag}%
  \BibitemOpen
  \bibfield  {author} {\bibinfo {author} {\bibfnamefont {A.}~\bibnamefont
  {Aboubrahim}}, \bibinfo {author} {\bibfnamefont {M.}~\bibnamefont {Klasen}},
  \ and\ \bibinfo {author} {\bibfnamefont {L.~P.}\ \bibnamefont {Wiggering}},\
  }\href {\doibase 10.1088/1475-7516/2023/08/075} {\bibfield  {journal}
  {\bibinfo  {journal} {JCAP}\ }\textbf {\bibinfo {volume} {08}},\ \bibinfo
  {pages} {075} (\bibinfo {year} {2023})},\ \Eprint
  {http://arxiv.org/abs/2306.07753} {2306.07753 [hep-ph]} \BibitemShut
  {NoStop}%
\bibitem [{\citenamefont {Bringmann}\ and\ \citenamefont
  {Hofmann}(2007)}]{Bringmann:2006mu}%
  \BibitemOpen
  \bibfield  {author} {\bibinfo {author} {\bibfnamefont {T.}~\bibnamefont
  {Bringmann}}\ and\ \bibinfo {author} {\bibfnamefont {S.}~\bibnamefont
  {Hofmann}},\ }\href {\doibase 10.1088/1475-7516/2007/04/016} {\bibfield
  {journal} {\bibinfo  {journal} {JCAP}\ }\textbf {\bibinfo {volume} {04}},\
  \bibinfo {pages} {016} (\bibinfo {year} {2007})},\ \bibinfo {note} {[erratum:
  JCAP \textbf{03} (2016), E02]},\ \Eprint
  {http://arxiv.org/abs/hep-ph/0612238} {hep-ph/0612238 [hep-ph]} \BibitemShut
  {NoStop}%
\bibitem [{\citenamefont {Sakharov}(1967)}]{Sakharov:1967dj}%
  \BibitemOpen
  \bibfield  {author} {\bibinfo {author} {\bibfnamefont {A.~D.}\ \bibnamefont
  {Sakharov}},\ }\href {\doibase 10.1070/PU1991v034n05ABEH002497} {\bibfield
  {journal} {\bibinfo  {journal} {Pisma Zh. Eksp. Teor. Fiz.}\ }\textbf
  {\bibinfo {volume} {5}},\ \bibinfo {pages} {32} (\bibinfo {year}
  {1967})}\BibitemShut {NoStop}%
\bibitem [{\citenamefont {Witten}(1982)}]{Witten:1982fp}%
  \BibitemOpen
  \bibfield  {author} {\bibinfo {author} {\bibfnamefont {E.}~\bibnamefont
  {Witten}},\ }\href {\doibase 10.1016/0370-2693(82)90728-6} {\bibfield
  {journal} {\bibinfo  {journal} {Phys. Lett. B}\ }\textbf {\bibinfo {volume}
  {117}},\ \bibinfo {pages} {324} (\bibinfo {year} {1982})}\BibitemShut
  {NoStop}%
\bibitem [{\citenamefont {Brust}\ \emph {et~al.}(2013)\citenamefont {Brust},
  \citenamefont {Kaplan},\ and\ \citenamefont {Walters}}]{Brust:2013xla}%
  \BibitemOpen
  \bibfield  {author} {\bibinfo {author} {\bibfnamefont {C.}~\bibnamefont
  {Brust}}, \bibinfo {author} {\bibfnamefont {D.~E.}\ \bibnamefont {Kaplan}}, \
  and\ \bibinfo {author} {\bibfnamefont {M.~T.}\ \bibnamefont {Walters}},\
  }\href {\doibase 10.1007/JHEP12(2013)058} {\bibfield  {journal} {\bibinfo
  {journal} {JHEP}\ }\textbf {\bibinfo {volume} {12}},\ \bibinfo {pages} {058}
  (\bibinfo {year} {2013})},\ \Eprint {http://arxiv.org/abs/1303.5379}
  {1303.5379 [hep-ph]} \BibitemShut {NoStop}%
\bibitem [{\citenamefont {Steigman}(2012)}]{Steigman:2012ve}%
  \BibitemOpen
  \bibfield  {author} {\bibinfo {author} {\bibfnamefont {G.}~\bibnamefont
  {Steigman}},\ }\href {\doibase 10.1155/2012/268321} {\bibfield  {journal}
  {\bibinfo  {journal} {Adv. High Energy Phys.}\ }\textbf {\bibinfo {volume}
  {2012}},\ \bibinfo {pages} {268321} (\bibinfo {year} {2012})},\ \Eprint
  {http://arxiv.org/abs/1208.0032} {1208.0032 [hep-ph]} \BibitemShut {NoStop}%
\bibitem [{\citenamefont {Kolb}(1990)}]{Kolb:1990vq}%
  \BibitemOpen
  \bibfield  {author} {\bibinfo {author} {\bibfnamefont {E.~W.}\ \bibnamefont
  {Kolb}},\ }\href {\doibase 10.1201/9780429492860} {\bibfield  {journal}
  {\bibinfo  {journal} {Front. Phys.}\ }\textbf {\bibinfo {volume} {69}},\
  \bibinfo {pages} {1} (\bibinfo {year} {1990})}\BibitemShut {NoStop}%
\bibitem [{\citenamefont {Gondolo}\ and\ \citenamefont
  {Gelmini}(1991)}]{Gondolo:1990dk}%
  \BibitemOpen
  \bibfield  {author} {\bibinfo {author} {\bibfnamefont {P.}~\bibnamefont
  {Gondolo}}\ and\ \bibinfo {author} {\bibfnamefont {G.}~\bibnamefont
  {Gelmini}},\ }\href {\doibase 10.1016/0550-3213(91)90438-4} {\bibfield
  {journal} {\bibinfo  {journal} {Nucl. Phys. B}\ }\textbf {\bibinfo {volume}
  {360}},\ \bibinfo {pages} {145} (\bibinfo {year} {1991})}\BibitemShut
  {NoStop}%
\bibitem [{\citenamefont {Gondolo}\ \emph {et~al.}(2012)\citenamefont
  {Gondolo}, \citenamefont {Hisano},\ and\ \citenamefont
  {Kadota}}]{Gondolo:2012vh}%
  \BibitemOpen
  \bibfield  {author} {\bibinfo {author} {\bibfnamefont {P.}~\bibnamefont
  {Gondolo}}, \bibinfo {author} {\bibfnamefont {J.}~\bibnamefont {Hisano}}, \
  and\ \bibinfo {author} {\bibfnamefont {K.}~\bibnamefont {Kadota}},\ }\href
  {\doibase 10.1103/PhysRevD.86.083523} {\bibfield  {journal} {\bibinfo
  {journal} {Phys. Rev. D}\ }\textbf {\bibinfo {volume} {86}},\ \bibinfo
  {pages} {083523} (\bibinfo {year} {2012})},\ \Eprint
  {http://arxiv.org/abs/1205.1914} {1205.1914 [hep-ph]} \BibitemShut {NoStop}%
\bibitem [{\citenamefont {Visinelli}\ and\ \citenamefont
  {Gondolo}(2015)}]{Visinelli:2015eka}%
  \BibitemOpen
  \bibfield  {author} {\bibinfo {author} {\bibfnamefont {L.}~\bibnamefont
  {Visinelli}}\ and\ \bibinfo {author} {\bibfnamefont {P.}~\bibnamefont
  {Gondolo}},\ }\href {\doibase 10.1103/PhysRevD.91.083526} {\bibfield
  {journal} {\bibinfo  {journal} {Phys. Rev. D}\ }\textbf {\bibinfo {volume}
  {91}},\ \bibinfo {pages} {083526} (\bibinfo {year} {2015})},\ \Eprint
  {http://arxiv.org/abs/1501.02233} {1501.02233 [astro-ph.CO]} \BibitemShut
  {NoStop}%
\bibitem [{\citenamefont {Plumacher}(1997)}]{Plumacher:1996kc}%
  \BibitemOpen
  \bibfield  {author} {\bibinfo {author} {\bibfnamefont {M.}~\bibnamefont
  {Plumacher}},\ }\href {\doibase 10.1007/s002880050418} {\bibfield  {journal}
  {\bibinfo  {journal} {Z. Phys. C}\ }\textbf {\bibinfo {volume} {74}},\
  \bibinfo {pages} {549} (\bibinfo {year} {1997})},\ \Eprint
  {http://arxiv.org/abs/hep-ph/9604229} {hep-ph/9604229 [hep-ph]} \BibitemShut
  {NoStop}%
\bibitem [{\citenamefont {Harvey}\ and\ \citenamefont
  {Turner}(1990)}]{Harvey:1990qw}%
  \BibitemOpen
  \bibfield  {author} {\bibinfo {author} {\bibfnamefont {J.~A.}\ \bibnamefont
  {Harvey}}\ and\ \bibinfo {author} {\bibfnamefont {M.~S.}\ \bibnamefont
  {Turner}},\ }\href {\doibase 10.1103/PhysRevD.42.3344} {\bibfield  {journal}
  {\bibinfo  {journal} {Phys. Rev. D}\ }\textbf {\bibinfo {volume} {42}},\
  \bibinfo {pages} {3344} (\bibinfo {year} {1990})}\BibitemShut {NoStop}%
\bibitem [{\citenamefont {Nardi}\ \emph {et~al.}(2006)\citenamefont {Nardi},
  \citenamefont {Nir}, \citenamefont {Roulet},\ and\ \citenamefont
  {Racker}}]{Nardi:2006fx}%
  \BibitemOpen
  \bibfield  {author} {\bibinfo {author} {\bibfnamefont {E.}~\bibnamefont
  {Nardi}}, \bibinfo {author} {\bibfnamefont {Y.}~\bibnamefont {Nir}}, \bibinfo
  {author} {\bibfnamefont {E.}~\bibnamefont {Roulet}}, \ and\ \bibinfo {author}
  {\bibfnamefont {J.}~\bibnamefont {Racker}},\ }\href {\doibase
  10.1088/1126-6708/2006/01/164} {\bibfield  {journal} {\bibinfo  {journal}
  {JHEP}\ }\textbf {\bibinfo {volume} {01}},\ \bibinfo {pages} {164} (\bibinfo
  {year} {2006})},\ \Eprint {http://arxiv.org/abs/hep-ph/0601084}
  {hep-ph/0601084 [hep-ph]} \BibitemShut {NoStop}%
\bibitem [{\citenamefont {Kuzmin}\ \emph {et~al.}(1985)\citenamefont {Kuzmin},
  \citenamefont {Rubakov},\ and\ \citenamefont {Shaposhnikov}}]{Kuzmin:1985mm}%
  \BibitemOpen
  \bibfield  {author} {\bibinfo {author} {\bibfnamefont {V.~A.}\ \bibnamefont
  {Kuzmin}}, \bibinfo {author} {\bibfnamefont {V.~A.}\ \bibnamefont {Rubakov}},
  \ and\ \bibinfo {author} {\bibfnamefont {M.~E.}\ \bibnamefont
  {Shaposhnikov}},\ }\href {\doibase 10.1016/0370-2693(85)91028-7} {\bibfield
  {journal} {\bibinfo  {journal} {Phys. Lett. B}\ }\textbf {\bibinfo {volume}
  {155}},\ \bibinfo {pages} {36} (\bibinfo {year} {1985})}\BibitemShut
  {NoStop}%
\bibitem [{\citenamefont {Berezinsky}\ \emph {et~al.}(2003)\citenamefont
  {Berezinsky}, \citenamefont {Dokuchaev},\ and\ \citenamefont
  {Eroshenko}}]{Berezinsky:2003vn}%
  \BibitemOpen
  \bibfield  {author} {\bibinfo {author} {\bibfnamefont {V.}~\bibnamefont
  {Berezinsky}}, \bibinfo {author} {\bibfnamefont {V.}~\bibnamefont
  {Dokuchaev}}, \ and\ \bibinfo {author} {\bibfnamefont {Y.}~\bibnamefont
  {Eroshenko}},\ }\href {\doibase 10.1103/PhysRevD.68.103003} {\bibfield
  {journal} {\bibinfo  {journal} {Phys. Rev. D}\ }\textbf {\bibinfo {volume}
  {68}},\ \bibinfo {pages} {103003} (\bibinfo {year} {2003})},\ \Eprint
  {http://arxiv.org/abs/astro-ph/0301551} {astro-ph/0301551 [astro-ph]}
  \BibitemShut {NoStop}%
\bibitem [{\citenamefont {l.~Chen}\ \emph {et~al.}(2001)\citenamefont
  {l.~Chen}, \citenamefont {Kamionkowski},\ and\ \citenamefont
  {m.~Zhang}}]{Chen:2001jz}%
  \BibitemOpen
  \bibfield  {author} {\bibinfo {author} {\bibfnamefont {X.}~\bibnamefont
  {l.~Chen}}, \bibinfo {author} {\bibfnamefont {M.}~\bibnamefont
  {Kamionkowski}}, \ and\ \bibinfo {author} {\bibfnamefont {X.}~\bibnamefont
  {m.~Zhang}},\ }\href {\doibase 10.1103/PhysRevD.64.021302} {\bibfield
  {journal} {\bibinfo  {journal} {Phys. Rev. D}\ }\textbf {\bibinfo {volume}
  {64}},\ \bibinfo {pages} {021302} (\bibinfo {year} {2001})},\ \Eprint
  {http://arxiv.org/abs/astro-ph/0103452} {astro-ph/0103452 [astro-ph]}
  \BibitemShut {NoStop}%
\bibitem [{\citenamefont {Bhattacharya}\ \emph {et~al.}(2016)\citenamefont
  {Bhattacharya}, \citenamefont {Sahoo},\ and\ \citenamefont
  {Sahu}}]{Bhattacharya:2015qpa}%
  \BibitemOpen
  \bibfield  {author} {\bibinfo {author} {\bibfnamefont {S.}~\bibnamefont
  {Bhattacharya}}, \bibinfo {author} {\bibfnamefont {N.}~\bibnamefont {Sahoo}},
  \ and\ \bibinfo {author} {\bibfnamefont {N.}~\bibnamefont {Sahu}},\ }\href
  {\doibase 10.1103/PhysRevD.93.115040} {\bibfield  {journal} {\bibinfo
  {journal} {Phys. Rev. D}\ }\textbf {\bibinfo {volume} {93}},\ \bibinfo
  {pages} {115040} (\bibinfo {year} {2016})},\ \Eprint
  {http://arxiv.org/abs/1510.02760} {1510.02760 [hep-ph]} \BibitemShut
  {NoStop}%
\bibitem [{\citenamefont {{ATLAS
  Collaboration}}(2026)}]{ATLAS:2026disappearing}%
  \BibitemOpen
  \bibfield  {author} {\bibinfo {author} {\bibnamefont {{ATLAS
  Collaboration}}},\ }\href {\doibase 10.1007/JHEP07(2026)152} {\bibfield
  {journal} {\bibinfo  {journal} {JHEP}\ }\textbf {\bibinfo {volume} {07}},\
  \bibinfo {pages} {152} (\bibinfo {year} {2026})},\ \Eprint
  {http://arxiv.org/abs/2603.08315} {2603.08315 [hep-ex]} \BibitemShut
  {NoStop}%
\bibitem [{\citenamefont {Goldstein}\ and\ \citenamefont
  {Hill}(2026)}]{Goldstein:2026}%
  \BibitemOpen
  \bibfield  {author} {\bibinfo {author} {\bibfnamefont {S.}~\bibnamefont
  {Goldstein}}\ and\ \bibinfo {author} {\bibfnamefont {J.~C.}\ \bibnamefont
  {Hill}},\ }\href {\doibase 10.1103/5vk8-v1rb} {\bibfield  {journal} {\bibinfo
   {journal} {Phys. Rev. D}\ }\textbf {\bibinfo {volume} {114}},\ \bibinfo
  {pages} {L021305} (\bibinfo {year} {2026})},\ \Eprint
  {http://arxiv.org/abs/2603.13226} {2603.13226 [astro-ph.CO]} \BibitemShut
  {NoStop}%
\bibitem [{\citenamefont {Bergstrom}\ \emph {et~al.}(1998)\citenamefont
  {Bergstrom}, \citenamefont {Ullio},\ and\ \citenamefont
  {Buckley}}]{Bergstrom:1997fj}%
  \BibitemOpen
  \bibfield  {author} {\bibinfo {author} {\bibfnamefont {L.}~\bibnamefont
  {Bergstrom}}, \bibinfo {author} {\bibfnamefont {P.}~\bibnamefont {Ullio}}, \
  and\ \bibinfo {author} {\bibfnamefont {J.~H.}\ \bibnamefont {Buckley}},\
  }\href {\doibase 10.1016/S0927-6505(98)00015-2} {\bibfield  {journal}
  {\bibinfo  {journal} {Astropart. Phys.}\ }\textbf {\bibinfo {volume} {9}},\
  \bibinfo {pages} {137} (\bibinfo {year} {1998})},\ \Eprint
  {http://arxiv.org/abs/astro-ph/9712318} {astro-ph/9712318 [astro-ph]}
  \BibitemShut {NoStop}%
\bibitem [{\citenamefont {Cirelli}\ \emph {et~al.}(2011)\citenamefont
  {Cirelli}, \citenamefont {Corcella}, \citenamefont {Hektor}, \citenamefont
  {Hutsi}, \citenamefont {Kadastik}, \citenamefont {Panci}, \citenamefont
  {Raidal}, \citenamefont {Sala},\ and\ \citenamefont
  {Strumia}}]{Cirelli:2010xx}%
  \BibitemOpen
  \bibfield  {author} {\bibinfo {author} {\bibfnamefont {M.}~\bibnamefont
  {Cirelli}}, \bibinfo {author} {\bibfnamefont {G.}~\bibnamefont {Corcella}},
  \bibinfo {author} {\bibfnamefont {A.}~\bibnamefont {Hektor}}, \bibinfo
  {author} {\bibfnamefont {G.}~\bibnamefont {Hutsi}}, \bibinfo {author}
  {\bibfnamefont {M.}~\bibnamefont {Kadastik}}, \bibinfo {author}
  {\bibfnamefont {P.}~\bibnamefont {Panci}}, \bibinfo {author} {\bibfnamefont
  {M.}~\bibnamefont {Raidal}}, \bibinfo {author} {\bibfnamefont
  {F.}~\bibnamefont {Sala}}, \ and\ \bibinfo {author} {\bibfnamefont
  {A.}~\bibnamefont {Strumia}},\ }\href {\doibase
  10.1088/1475-7516/2012/10/E01} {\bibfield  {journal} {\bibinfo  {journal}
  {JCAP}\ }\textbf {\bibinfo {volume} {03}},\ \bibinfo {pages} {051} (\bibinfo
  {year} {2011})},\ \bibinfo {note} {[erratum: JCAP \textbf{10} (2012), E01]},\
  \Eprint {http://arxiv.org/abs/1012.4515} {1012.4515 [hep-ph]} \BibitemShut
  {NoStop}%
\bibitem [{\citenamefont {McDaniel}\ \emph {et~al.}(2024)\citenamefont
  {McDaniel}, \citenamefont {Ajello}, \citenamefont {Karwin}, \citenamefont
  {Mauro}, \citenamefont {Drlica-Wagner},\ and\ \citenamefont
  {S{\'a}nchez-Conde}}]{McDaniel:2023bju}%
  \BibitemOpen
  \bibfield  {author} {\bibinfo {author} {\bibfnamefont {A.}~\bibnamefont
  {McDaniel}}, \bibinfo {author} {\bibfnamefont {M.}~\bibnamefont {Ajello}},
  \bibinfo {author} {\bibfnamefont {C.~M.}\ \bibnamefont {Karwin}}, \bibinfo
  {author} {\bibfnamefont {M.~D.}\ \bibnamefont {Mauro}}, \bibinfo {author}
  {\bibfnamefont {A.}~\bibnamefont {Drlica-Wagner}}, \ and\ \bibinfo {author}
  {\bibfnamefont {M.~A.}\ \bibnamefont {S{\'a}nchez-Conde}},\ }\href {\doibase
  10.1103/PhysRevD.109.063024} {\bibfield  {journal} {\bibinfo  {journal}
  {Phys. Rev. D}\ }\textbf {\bibinfo {volume} {109}},\ \bibinfo {pages}
  {063024} (\bibinfo {year} {2024})},\ \Eprint
  {http://arxiv.org/abs/2311.04982} {2311.04982 [astro-ph.HE]} \BibitemShut
  {NoStop}%
\bibitem [{\citenamefont {Abdalla}\ and\ \citenamefont {{H.E.S.S.
  Collaboration}}(2022)}]{HESS:2022ygk}%
  \BibitemOpen
  \bibfield  {author} {\bibinfo {author} {\bibfnamefont {H.}~\bibnamefont
  {Abdalla}}\ and\ \bibinfo {author} {\bibnamefont {{H.E.S.S.
  Collaboration}}},\ }\href {\doibase 10.1103/PhysRevLett.129.111101}
  {\bibfield  {journal} {\bibinfo  {journal} {Phys. Rev. Lett.}\ }\textbf
  {\bibinfo {volume} {129}},\ \bibinfo {pages} {111101} (\bibinfo {year}
  {2022})},\ \Eprint {http://arxiv.org/abs/2207.10471} {2207.10471
  [astro-ph.HE]} \BibitemShut {NoStop}%
\bibitem [{\citenamefont {Acharyya}\ and\ \citenamefont {{CTA
  Collaboration}}(2021)}]{CTA:2020qlo}%
  \BibitemOpen
  \bibfield  {author} {\bibinfo {author} {\bibfnamefont {A.}~\bibnamefont
  {Acharyya}}\ and\ \bibinfo {author} {\bibnamefont {{CTA Collaboration}}},\
  }\href {\doibase 10.1088/1475-7516/2021/01/057} {\bibfield  {journal}
  {\bibinfo  {journal} {JCAP}\ }\textbf {\bibinfo {volume} {01}},\ \bibinfo
  {pages} {057} (\bibinfo {year} {2021})},\ \Eprint
  {http://arxiv.org/abs/2007.16129} {2007.16129 [astro-ph.HE]} \BibitemShut
  {NoStop}%
\bibitem [{\citenamefont {Bierlich}\ \emph {et~al.}(2022)\citenamefont
  {Bierlich}, \citenamefont {Chakraborty}, \citenamefont {Desai}, \citenamefont
  {Gellersen}, \citenamefont {Helenius}, \citenamefont {Ilten}, \citenamefont
  {L{\"o}nnblad}, \citenamefont {Mrenna}, \citenamefont {Prestel},
  \citenamefont {Preuss}, \citenamefont {Sj{\"o}strand}, \citenamefont
  {Skands}, \citenamefont {Utheim},\ and\ \citenamefont
  {Verheyen}}]{Bierlich:2022pythia83}%
  \BibitemOpen
  \bibfield  {author} {\bibinfo {author} {\bibfnamefont {C.}~\bibnamefont
  {Bierlich}}, \bibinfo {author} {\bibfnamefont {S.}~\bibnamefont
  {Chakraborty}}, \bibinfo {author} {\bibfnamefont {N.}~\bibnamefont {Desai}},
  \bibinfo {author} {\bibfnamefont {L.}~\bibnamefont {Gellersen}}, \bibinfo
  {author} {\bibfnamefont {I.}~\bibnamefont {Helenius}}, \bibinfo {author}
  {\bibfnamefont {P.}~\bibnamefont {Ilten}}, \bibinfo {author} {\bibfnamefont
  {L.}~\bibnamefont {L{\"o}nnblad}}, \bibinfo {author} {\bibfnamefont
  {S.}~\bibnamefont {Mrenna}}, \bibinfo {author} {\bibfnamefont
  {S.}~\bibnamefont {Prestel}}, \bibinfo {author} {\bibfnamefont {C.~T.}\
  \bibnamefont {Preuss}}, \bibinfo {author} {\bibfnamefont {T.}~\bibnamefont
  {Sj{\"o}strand}}, \bibinfo {author} {\bibfnamefont {P.}~\bibnamefont
  {Skands}}, \bibinfo {author} {\bibfnamefont {M.}~\bibnamefont {Utheim}}, \
  and\ \bibinfo {author} {\bibfnamefont {R.}~\bibnamefont {Verheyen}},\ }\href
  {\doibase 10.21468/SciPostPhysCodeb.8} {\bibfield  {journal} {\bibinfo
  {journal} {SciPost Phys. Codeb.}\ }\textbf {\bibinfo {volume} {2022}},\
  \bibinfo {pages} {8} (\bibinfo {year} {2022})},\ \Eprint
  {http://arxiv.org/abs/2203.11601} {2203.11601 [hep-ph]} \BibitemShut
  {NoStop}%
\end{thebibliography}%

\end{document}